\documentclass{emulateapj}
\usepackage{epstopdf}

\slugcomment{Accepted by the Astronomical Journal}

\shorttitle{BRAVA Data Release}
\shortauthors{Kunder et al.}

\begin{document}

\title{The Bulge Radial Velocity Assay ({\sl BRAVA}): II.  Complete Sample and Data Release}

\author{Andrea Kunder\altaffilmark{1},
Andreas Koch\altaffilmark{2}, 
R. Michael Rich\altaffilmark{3},
Roberto de Propris\altaffilmark{1},  
Christian D. Howard\altaffilmark{4},
Scott A. Stubbs\altaffilmark{1}, 
Christian I. Johnson\altaffilmark{3,5},
Juntai Shen\altaffilmark{6},
Yougang Wang\altaffilmark{7},
Annie C. Robin\altaffilmark{8}
John Kormendy\altaffilmark{9},
Mario Soto\altaffilmark{10},
Peter Frinchaboy\altaffilmark{11},
David B. Reitzel\altaffilmark{12,3},
HongSheng Zhao\altaffilmark{13},
\and
Livia Origlia\altaffilmark{14}
}

\altaffiltext{1}{Cerro Tololo Inter-American Observatory, Casilla 603, La Serena, Chile}
\affil{E-mail: akunder@ctio.noao.edu}
\altaffiltext{2}{Zentrum f\"ur Astronomie der Universit\"at Heidelberg, Landessternwarte, K\"onigstuhl 12, Heidelberg, Germany}
\altaffiltext{3}{Department of Physics and Astronomy, University of California, Los Angeles, CA 90095-1562}
\altaffiltext{4}{NASA Ames Research Center, MS 211-3 Moffett Field, CA 94035}
\altaffiltext{5}{National Science Foundation Astronomy and Astrophysics Postdoctoral Fellow}
\altaffiltext{6}{Key Laboratory for Research in Galaxies and Cosmology, Shanghai Astronomical Observatory, Chinese Academy of Sciences, Shanghai 200030, China}
\altaffiltext{7}{Key Laboratory of Optical Astronomy, National Astronomical Observatoires, Chinese Academy of Sciences, Beijing 100012, China}
\altaffiltext{8}{Observatoire des sciences de l'Univers de Besan\c{c}on, Besan\c{c}on, France}
\altaffiltext{9}{Department of Astronomy, The University of Texas at Austin, Austin, TX 78712}
\altaffiltext{10} {Universidad de la Serena, Benavente 980 Casilla 544, La Serena ,  Chile}
\altaffiltext{11} {Texas Christian University, Department of Physics and Astronomy, Fort Worth, TX 76129}
\altaffiltext{12}{Griffith Observatory, 2800 East Observatory Road, Los Angeles, CA 90027, USA}
\altaffiltext{13}{SUPA, School of Physics and Astronomy, University of St. Andrews, KY16 9SS, UK}
\altaffiltext{14}{INAF, Osservatorio Astronomico di Bologna, Via Ranzani 1, IT 40127 Bologna, Italy}

\begin{abstract}
We present new radial velocity measurements from the Bulge Radial Velocity Assay ({\sl BRAVA}), 
a large scale spectroscopic survey of M-type giants in the Galactic bulge/bar region.   The sample 
of $\sim$4500 new radial velocities, mostly in the region $-10^\circ < l < +10^\circ$ and 
$b\approx -6^{\circ}$ more than doubles the existent published data set.   Our new data extend 
our rotation curve and velocity dispersion profile to $+20^\circ$, which is $\sim$2.8 kpc from the 
Galactic Center.  The new data confirm the cylindrical rotation observed at $-6^\circ$ and 
$-8^\circ$, and are an excellent fit to the Shen et al. (2010) N-body bar model.  We 
measure the strength of the TiO$\varepsilon$ molecular band as a first step towards a 
metallicity ranking of the stellar sample, from which we confirm the presence of a vertical 
abundance gradient.  Our survey finds no strong evidence of previously unknown 
kinematic streams.  We also publish our complete catalog of radial velocities, photometry, 
TiO band strengths, and spectra, which is available at the IRSA archive:
{\tt http://irsa.ipac.caltech.edu/} as well as at UCLA: {\tt http://brava.astro.ucla.edu/}.
\end{abstract}


\section{Introduction}

Only for stars in the Milky Way is it currently possible to examine both the three dimensional 
kinematics and composition of a central bulge/bar population, offering a unique laboratory for 
the study of galaxy formation and evolution.   Up until now, there have been very few optical 
spectra and radial velocities of bulge stars, published in a catalog form, with the spectra and 
measurements made publicly available.   Here we present a catalog of the {\it Bulge Radial 
Velocity Assay}  (BRAVA) that has a sample of $\sim 10,000$ M giant stars selected from the 
red giant branch (RGB) of the Two micron All-sky Survey \citep[2MASS;][]{skrutskie06}.  This 
catalog gives the 2MASS magnitudes and positions, our measured radial velocities, and our 
spectra.  This publication presents our complete sample of low resolution spectra for the 
{\sl BRAVA} survey, which covers approximately $-10^\circ < l <+10^\circ $ 
and $-4^\circ <b < -8^\circ $.  We 
also present a rich new data set at $b=-6^\circ$ that complements the first results presented in 
Howard et~al. (2008, hereafter Paper I) and the properties of which conform to our other 
latitudinal studies at $-4^\circ$ and $-8^\circ$.  We expect that the database will ultimately be 
useful in constraining the dynamical model of the Galactic  bulge, and in placing limits on the 
detection of cold streams from fossil infall events.  The 2MASS astrometry is of sufficient quality 
that proper motions for our sample will be measurable in the next few years.  

Although evidence for a barred potential has long been known \citep[e.g.][]{liszt80}, the modeling 
of the 2 micron surface brightness distribution in the bulge as a bar \citep{blitz91} 
established that the stellar distribution of the bulge is best modeled as a bar with the
position angle of its major axis as $20-45^\circ$.  The self-consistent dynamical model 
of a rapidly rotating bar \citep{zhao96} was 
a significant advance, and a number of models have followed, however, additional progress 
requires constraining new models with a much larger kinematic data set.  The question of exactly 
how the bar structure formed -- presumably via some kind of secular evolution \citep{kormendy04}, 
but including rapid star formation, winds, and dissipation -- can be better addressed with these new 
large samples.  Additionally, the relationship between the bulge, inner disk, thick disk and halo, may be 
illuminated further.  As the coverage and sample sizes analyzed in the bulge increase, it becomes
possible to place meaningful limits on the influence of infalling satellite and other types of merger 
events.  The ultimate goal is to move beyond the characterization of the bar's morphology and orientation to 
the characterization of the bulge/bar as a stellar system with a unique and complex history.

Early surveys of the bulge \citep{nassau58} immediately revealed a distinguishing 
characteristic, namely large numbers of M giants compared with the giant population 
in globular clusters and the halo.  The breakthrough in achieving a physical understanding 
of the M giant population arose from the surveys undertaken by Victor and Betty Blanco 
using the newly commissioned grating/prism on the 4m telescope at CTIO 
(Blanco, McCarthy, \& Blanco 1984) -- an effort that yielded thousands of photographic low 
resolution classification spectra.  These studies provided the raw material that fed the first 
radial velocity study \citep{mould86} and infrared surveys \citep[e.g.][]{frogel87}.  The cool bulge 
M giants could be easily identified and exploited because of their brightness in the near IR 
(where extinction is low) and by their strong molecular 
(chiefly titanium oxide $\rm[TiO]$) bands.  Sharples, Walker \& Cropper (1990) used an early fiber 
spectrograph on the Anglo-Australian Telescope to obtain $\sim 300$ spectra of bulge M giants 
in Baade's Window.  However, no study specifically exploited the  M giants as kinematic probes 
over the entire bulge, even though they could be easily identified from their extremely red $V-I$ colors.

Following the self consistent, rapidly rotating bar model of \citep{zhao96}, two efforts modeled newly 
available, substantial data sets.  \citet{sevenster99} employed the OH/IR star population to 
constrain a bar model, while the \citet{beaulieu00} study used planetary nebulae (PNe) as 
kinematic probes of the bulge and fit a range of bar models to the data.   Both probes offered an 
advantage in that they are ubiquitous and easy to identify over the whole of the bulge.   In the case 
of the PNe, the numbers are relatively few (373).  Unfortunately, PNe are not specific to a given 
stellar population so disk contamination is a possibility, and the total number in the PNe sample 
is modest.  However, \citet{beaulieu00} were able to test both self-consistent and N-body 
bar models against the PNe kinematic data.  \citet{sevenster99} were able to draw significant 
constraints from the OH/IR star kinematics, finding a bar-shaped bulge at a roughly 45$^\circ$ bar 
angle.  The OH/IR stars are also rare, with numbers similar to those of the PNe.  
 
With the release of the 2MASS database, it became possible to easily select M giants over the 
entirety of the Galactic bulge.  The 2MASS survey offered an essentially unlimited supply of 
kinematic probes.  The {\sl Bulge Radial Velocity Assay} or {\sl BRAVA}  (see also Rich et al. 
2007b and Paper I) was conceived to exploit the high quality, uniform, photometry and astrometry 
for this data set.    We select red giants from the K, J$-$K red giant branch that are approximately 
at the distance to the Galactic center, R$_{\rm 0}$=8 kpc, and are highly likely to be 
bulge/bar members.  These stars can thus be selected from highly obscured regions of the 
bulge.  The first results of the {\sl BRAVA} project  are given additionally in Rich et al. (2007b), Paper I, 
Howard et al. (2009), and Shen et al. (2010).  
The principle result, based on $\sim$4500 stars, was the confirmation of cylindrical rotation,
and that the simple bar/boxy bulge model matches
the BRAVA kinematics strikingly well with no need for a merger-made
classical bulge \citep{shen10}.  The 
sample of red giant kinematic probes is now 8585.   We also have in progress a new 
version of the self-consistent model (Wang et al. 2011).

As described in \citet{rich07b}, we used the Hydra spectrograph on the Blanco 4m telescope, 
set at a central wavelength of $\sim$7900\AA \footnote{The wavelength range for 
each {\sl BRAVA} field
is given in our data release, as it varies slightly depending on what year the observations
were taken.}.  Many details of the observation and analysis are 
given in Paper I and will not be repeated here.  After some experimentation, we ultimately determined 
that even with the prominent, partially overlapping, molecular band absorption, the near-infrared 
Ca triplet (CaT) offered the best opportunity to obtain excellent radial velocity cross correlation.

While originally conceived as a purely kinematic study, the data quality of {\sl BRAVA} allows us, 
in principle, to investigate the metallicity distribution of the bulge.  Only with the addition of chemical 
abundance information will we ultimately be able to pin down the formation mechanisms that laid the 
kinematic traces found in our survey.  In particular, the presence of a radial or vertical metallicity gradients in 
the bulge may be indicative formation mechanisms \citep{melendez08, zoccali08, babusiaux10, johnson11} 
and warrants a well calibrated, large-number sample, such as {\sl BRAVA}. 
Here we report the coordinates, photometry,  TiO molecular band strengths, and radial velocities of 
individual red giants in the {\sl BRAVA} fields, as well as the mean velocities and velocity dispersions 
for each field.  The data release also includes all of our actual reduced spectra.  It is our intention to include proper motion data to the compilation, as well.
The present paper describes the final data set of {\sl BRAVA} and presents the {\sl BRAVA}
data release and website, which includes all available data and spectra from our observing campaign. 
We also refine the analysis presented in \cite{shen10} and confirm that the Galactic bulge appears
to consist of a massive bar undergoing pure cylindrical rotation and yield strong limits
on the presence of a classical bulge.

The structure of the paper is as follows. We present the observations and spectroscopy,
especially insofar as these were changed from the setups used by Paper I and \citet{howard09}
in \textsection1. Velocity calibrations and the main results are discussed in the next
section, \textsection2, where we also introduce our public data release.
Our molecular line index measurements are then introduced in \textsection3.  \textsection4
shows the new data fit to the \citet{shen10} N-body bar model, before 
concluding the data release in \textsection5. 

\section{New Observations and Spectroscopy}

The observations presented here were taken in 2008 with the Hydra multifiber bench 
spectrograph at the Cassegrain focus of the Blanco 4m telescope at the Cerro Tololo 
Inter-American Observatory (CTIO)\footnote{CTIO is operated by AURA, Inc., under contract
to the National Science Foundation.}.  Nineteen individual bulge fields were observed in July 
2008 and twenty-three in August 2008.  Additionally, three radial velocity standard stars 
(HD~177017, HD~218541 and HD~146051), already used in Paper I,  were acquired during 
these runs, as well as observations of the {\sl BRAVA} calibration field at $(l,b)=(6,-4)$. 

Howard et~al. (2008, hereafter Paper I) describe the observational setup and sample 
selection to obtain radial velocities of the bulge M giants.  Briefly, in 2005, our central 
wavelength was $\sim$7600\AA, but every year it was adjusted redward in order to 
observe the calcium triplet as well as the TiO band at $\sim$7050.  The specific
wavelength range for each {\sl BRAVA} field is listed in our data release.  For our 2008
observations, the KPGLD grating was employed, blazed at 8500\AA \ and with 1x2 binning.  
The central wavelength was $\sim$7900\AA \ and the effective dispersion was 0.88 \AA \ 
pixel$^{-1}$ with an effective resolving power of $R \sim 4200$.  
The spectral range of the new data set included all three CaT lines, and for
each field, three exposures at 600 s each were obtained.  These fields typically allowed for 
$\sim$106 fibers to be place on M giants and an additional $\sim$20 fibers to be used for sky 
subtraction.  The signal-to-noise ($S/N$) at $\sim$7500 \AA \ ranges from 10-80, with median
values of $\sim$ 35-45.  The variations in S/N arise mainly from transparency at the time the 
data were taken, and the field position in the bulge (extinction, crowding).  

\begin{figure}[htb]  
\includegraphics[height=0.67\hsize]{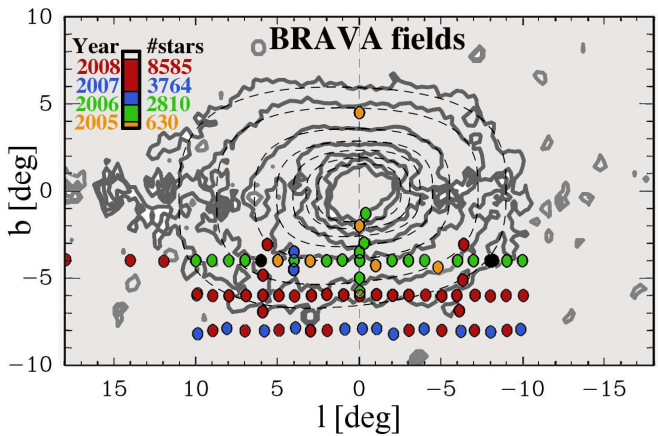}
\caption{All the observed {\sl BRAVA} fields, overplotted on the COBE 2 $\micron$ image (Launhardt, 
Zylka, \& Mezger 2002).  Each field is 40' and are color coded to designate the year observed.  The 
black circles are fields observed over multiple years. 
\label{BRflds}}
\end{figure}

A complete listing of the observed fields is shown in Table~\ref{obs}, and
all the observed fields from Paper I as well as those presented here are shown in 
Figure~\ref{BRflds}.

\begin{figure}[htb]
\begin{center}
\includegraphics[width=1\hsize]{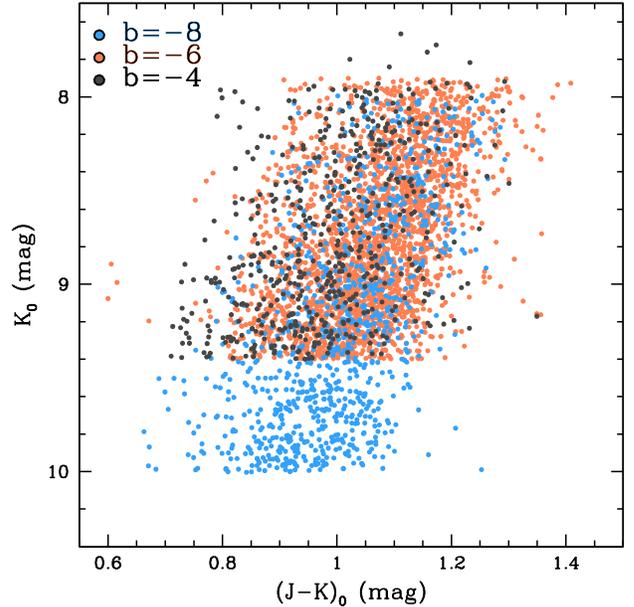}
\end{center}
\caption{The dereddened 2MASS color-magnitude diagram of the new 4352 M giants 
for which radial velocities have been determined.  Their position on the CMD indicates
they are bona fide bulge M giants.
\label{cmd}}
\end{figure}

The M giants were selected from the {\sl 2MASS} catalog; hence
uniform and consistent astrometry and photometry is known for all targets.  Figure~\ref{cmd}
shows the color-magnitude diagram of the targets.  Here the reddening values are taken 
from \citet{schlegel98}.  Our tests showed that application of the
\citet{schlegel98} values led to a consistent decrease in the width of the RGB in lower latitude 
fields, and gave a good overal consistency between the various fields, which span a wide 
range in Galactic latitude and longitude.  Reddening and the
abundance of target stars decreases as a function of galactic latitude, so the fields at $b$=$-$8$^\circ$
reach to K$\sim$ 10.1 mag whereas the fields at $b$=$-$4$^\circ$ and $b$=$-$6$^\circ$ go to
K$\sim$9.5 mag.  This selection was shown in Paper I and by \citet{howard09}
to largely avoid the red clump in the bulge as well as the M giants belonging to the disk.

The data were reduced and the radial velocities were obtained in a similar 
manner as described in Paper I.  After trimming, overscan- and bias correcting 
the spectra, cosmic rays were removed using the Laplacian edge-detection
routine, L.A.Cosmic \citep{dokkum01}.  The IRAF task {\tt dohydra} was used for 
aperture identification and tracing, flat-fielding, wavelength calibration and
sky subtraction.  The cross-correlation routine, {\tt xcsao}, in IRAF was then used to 
cross-correlate the spectra against our three radial velocity standard stars,   
obtain the radial velocities and correct them to the heliocentric rest frame.  The 
radial velocity standard stars were also used to apply the appropriate zero-point 
shift of the velocities.

\begin{figure}[htb]  
\includegraphics[width=1\hsize]{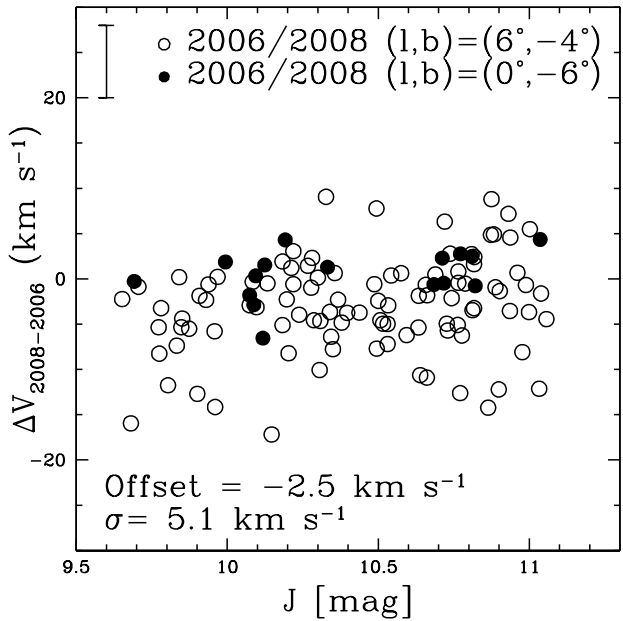}
\caption{The offset in stellar radial velocities between Paper I observations and the 
2008 observations for field $(l,b)=(6,-4)$ (open circles) and for field $(l,b)=(0,-6)$ (closed circles).  
The size of the errors in
the 2008 velocities as reported by $xcsao$ is shown in the top left corner.
\label{comp0608a}}
\end{figure}

To assess the consistency of our velocity results, a star-by-star comparison of 
stellar velocities was conducted of the field at $(l,b)=(6,-4)$ as well as overlap stars from
the field at $(l,b)=(0,-6)$.  Figure~\ref{comp0608a}
shows 101 stellar velocities in field $(l,b)=(6,-4)$ and the 16 velocities in field 
at $(l,b)=(0,-6)$ that are in common between the 2006 and 2008 data.
As these velocities agree to within 5 km s$^{-1}$,  (the 1$\sigma$ dispersion of the difference), 
this is the value adopted as the global, individual stellar velocity error.  This error is identical 
to that found for the radial velocities in Paper I.  This error is also similar to the
velocity precision obtained with our spectra as reported by {\tt xcsao}.

The heliocentric velocities are corrected for
the reflex motion of the Sun following \citet{beaulieu00} and the velocity dispersion 
is given by
\begin{equation}
\sigma^{2}_{intrinsic}=\sigma^{2}_{observed}-\sum_{i=0}^{N}[error^{2}(v_{z})]/[2(N-1)], 
\end{equation}
where $\sigma^2_{observed}$ is the observed velocity dispersion of a given field.  The second term 
represents the uncertainties in the individual stellar radial velocity ($v_{z}$) measurements, 
where $N$ is the number of stars in that field.  Given the large amount of stars in each field, the second term is negligible, and hence $\rm \sigma^{2}_{intrinsic}=\sigma^{2}_{observed}$.
The average heliocentric and galactocentric velocities of each field is listed in Table~\ref{obs}.
A 4 $\sigma$-clipping algorithm is used to obtain these values; only
one star is affected by this clipping.  

The complete {\sl BRAVA} catalog can be accessed at the Infrared Science Archive (IRSA) 
{\tt http://irsa.ipac.caltech.edu/ } as well as at the University of California, Los Angeles (UCLA):
{\tt http://brava.astro.ucla.edu/}.  Each {\sl BRAVA} field is designated by its Galactic $l$ and $b$,
followed by the average helocentric and galactocentric velocity, error in velocity, dispersion, error
in dispersion, as described above.  Lastly the wavelength range of the spectrum is given, as
well as the number of stars in each field. 
The catalog includes all data from 2005, 2006, 2007 and 2008.  Following the link beside each field 
allows the measurements of the individual stars in the field to be obtained, as well as the reduced 
spectra in fits format.  An example of the format for the table of the individual radial velocities is 
shown in Table~\ref{ind}.

\subsection{Color/Magnitude Bias}
\begin{figure}[htb]  
\includegraphics[width=1\hsize]{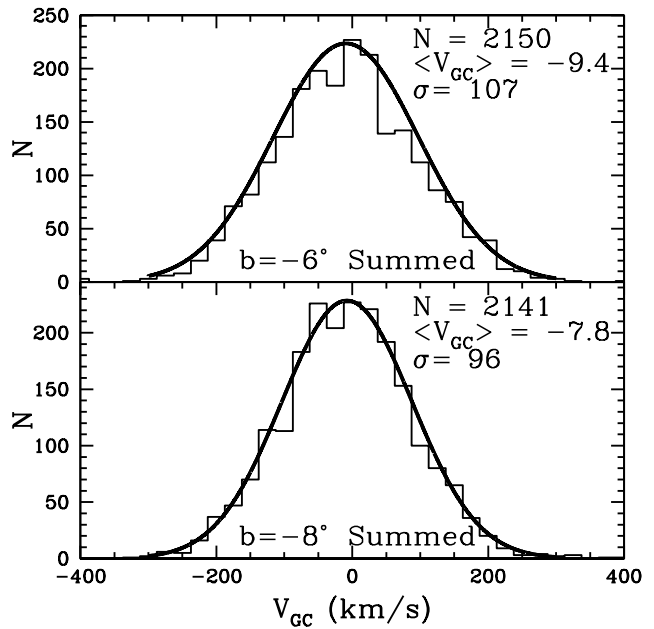}  
\caption{Histogram of all bulge RGB star galactocentric velocities for the fields at $b$=$-$8$^\circ$
(bottom) and $b$ = $-$6$^\circ$ (top).  Both distributions are normally distributed, with negligible skew
and kurtosis, which is consistent with a single kinematic population.
\label{bm68_strip}}
\end{figure}

It has been shown in Paper I and \citet{howard09} that bulge radial velocities summed over
minor- and major-axis fields consist of a Gaussian distribution with no apparent deviation from
a normal distribution.  The data analyzed in Paper I is located physically closer to the Galactic center
than the new data presented here.  As the latter is located closer
to the bulge/halo boundary, a comparison between the data presented here and that shown to be from a {\em bona fide} RGB bulge sample in Paper I can be used
to investigate the possibility of color and/or magnitude bias in the present sample. 
The bulge RGB stars from the $b$ = $-$8$^\circ$ and $b$ = $-$6$^\circ$ fields are summed and 
shown in Figure~\ref{bm68_strip}.  
They yield an apparent Gaussian distribution with $\rm <V_{GC}>$ = $-$8 $\pm$ 3 km s$^{-1}$ and 
$\sigma$ = 96 $\pm$ 2 km s$^{-1}$ for the $b$ = $-$8$^\circ$ fields and 
$\rm <V_{GC}>$ = $-$10 $\pm$ 3 km s$^{-1}$ and 
$\sigma$ = 107 $\pm$ 3 km s$^{-1}$ for the $b$ = $-$6$^\circ$ fields. 
Both of these curves are consistent not only with each other, but also with those from Paper I,
which suggests that these bulge samples in fact consist of a homogeneous, normally distributed 
stellar population.   Any subpopulation is not significant enough to cause a departure from 
the Gauassian distribution.

\section{TiO$\varepsilon$ as metallicity indicator}
While  our spectral range, in principle, offers a wealth of metallicity indicators, 
the standard calibrations of the line strengths of the three prominent \ion{Ca}{2} 
lines at 8498, 8542, 8662\AA~onto an [Fe/H] scale \citep[e.g.,][]{rutledge97} fail for 
the cool M-giants that make up the majority of the {\sl BRAVA} sample. 
Here, the largest problem for measuring the CaT is its contamination with the strong 
TiO$\varepsilon$ band at 8430\AA, which becomes progressively stronger with 
decreasing T$_{\rm eff}$.  In fact, this band has a strong dependence on stellar effective temperature \citep[e.g.][]{milone94}.   Since we can estimate our T$_{\rm eff}$-scale 
using the 2MASS infrared colors (Figure~\ref{jk_tio_mod}), measurements of the band 
strength then allows to explore the variations with stellar 
metallicities \citep{sharples90}.  In particular, at an expected mean of $\rm [Fe/H]$$\sim -0.5$ dex, 
the TiO$\varepsilon$ band will be very prominent in the metal rich bulge stars. 

As a first step towards a comprehensive metallicity distribution,
we follow \citep{sharples90} in extracting a TiO-strength 
index, TiO$\varepsilon$, from our spectra, which is defined as a magnitude via
\begin{equation}
{\rm TiO}\varepsilon \,=\, -2.5\log\,F_{1} / F_{2}
\end{equation}
where the fluxes are obtained from straight integration in the bands from 8370--8420 \AA~($F_1$) 
and 8440--8490 \AA~($F_2$).  The 1$\sigma$ error on this number, estimated from the 
variance in those spectral bands, is typically 0.05 mag.  Values of each star's TiO index 
are provided in the final {\sl BRAVA} catalogue\footnote{In a future work (Koch et al. in prep.) we
will explore the reliable calibration of the CaT line strengths onto metallicity for a subsample 
of the stars with warmer temperatures, in which the CaT is yet unaffected by TiO absorption. 
Our measured equivalent widths will then be added to this {\sl BRAVA} release.}.
Note that this molecular band is not measurable in the spectra taken in 2005 due to the 
instrumental set-up.

\begin{figure}[htb]
\begin{center}
\includegraphics[width=1\hsize]{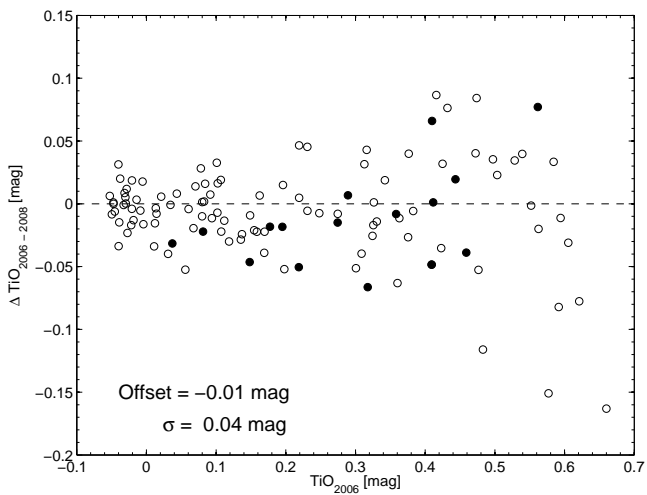}
\end{center}
\caption{
The offset in TiO$\varepsilon$ between stars observed in both 2006 and 2008. 
The filled circles correspond to stars in field $(l,b)=(0,-6)$, and the open circles 
correspond to stars in field $(l,b)=(6,-4)$.
\label{DeltaTiO}}
\end{figure}

\begin{figure}[htb]
\begin{center}
\includegraphics[width=1\hsize]{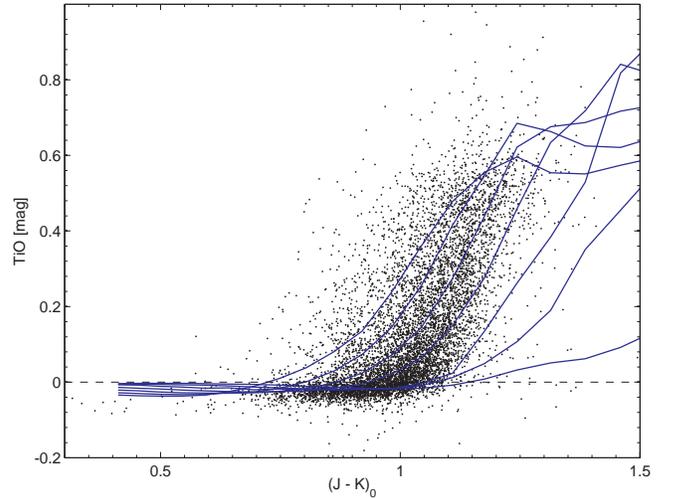}
\end{center}
\caption{TiO$\varepsilon$ index (after Sharples et al. 1990) versus infrared color. The solid lines 
in the bottom panel are based on $\alpha$-enhanced Phoenix models for metallicities, $\rm [M/H]$ 
between $-$2.5 and +0.5 dex (right to left) in steps of 0.5, adopting log\,$g$=1. 
A typical error bar is indicated in the top left corner.
\label{jk_tio_mod}}
\end{figure}

Figure~\ref{DeltaTiO} compares our TiO measurements for the two sets of overlapping 
spectra, as is already done for the velocities in Fig. 3.  The 1$\sigma$ scatter between both 
sets of TiO$\varepsilon$ is 0.04 mag with a mean difference 
below 0.01 mag. This is fully consistent with the uncertainty on the index, estimated above 
from the spectral variance.

In Figure~\ref{jk_tio_mod} the TiO$\varepsilon$ index is shown as a function of infrared color.
The distribution of the molecular index in our spectra shows a narrow band around zero, 
i.e., there is no discernible TiO depression, but a well-defined pseudo-continuum in the 
warmer stars. This holds for a broader range in J$-$K between 0.7 and 1.1 mag. On the 
other hand, for stars redder than J$-$K$\ga$1.0 (corresponding to T$_{\rm eff}$$\la$3700 K) 
there is a notable presence of progressively strong bands, leading to a linear rise in the TiO index. 
\subsection{Phoenix models}
A dispersion in TiO$\epsilon$ at fixed $\rm (J-K)$ can be due to both
differential reddening and metallicity spread.
To illustrate this, we consulted 
a set of synthetic spectra based on the Phoenix stellar atmosphere models. These offer 
a large grid of stellar atmospheric parameters and a spectral library in high 
resolution\footnote{\url{ftp://phoenix.hs.uni-hamburg.de/GAIA\_HighRes/Spectra/}} 
(Hauschildt, Allard \& Baron 1999, Hauschildt et~al. 2003).  In particular, its meticulous 
treatment of cool stellar atmospheres and its vast input sets of molecular opacities makes 
this synthetic library ideally suited for comparison with the {\sl BRAVA} data. 
Our chosen model grid consisted of T$_{\rm eff}$ ranging from 3200--4500 K in steps of 
100 K, log\,g\,=\,0, 0.5, 1.0, 1.5, metallicities [M/H] = $-2.5$ to $+0.5$ in increments of 
0.5 dex, and $\alpha$-enhancement of +0.4 dex as well as solar-scaled abundances, 
i.e., [$\alpha$/Fe]=0 dex. This generously comprises the range of parameters spanned 
by the {\sl BRAVA} late-type giant sample and is also consistent with the range in spectral types 
targeted in abundance studies of comparable stellar samples \citep{rich05, fulbright06, rich07a}.  
The synthetic spectra were then smoothed with a Gaussian kernel to mimic the resolution 
of the {\sl BRAVA} spectra. Finally, all the relevant quantities such as the TiO$\varepsilon$ index 
were derived from these convolved spectra in the identical manner as for the observed data. 

As Figure~\ref{jk_tio_mod} indicates, the bulk of the target stars falls near $\rm [M/H]$ of 
$-$0.5 dex, which is consistent with the metallicity distribution functions in the 
literature\footnote{Note, however, that such
observations strictly derive MDFs based on iron abundances, $\rm [Fe/H]$, and a
quantitative comparison with the models' global metallicity may not be
straighforward.} 
\citep[e.g.,][]{rangwala09, depropris11, johnson11}.  Moreover, stars are found to cover a 
broad range from super-solar down to low metallicities around $-1.5$ dex, with a hint of a 
very metal poor extension towards $-2$ dex. A detailed, quantitative study of the abundance 
properties of the {\sl BRAVA} sample will be presented in a forthcoming paper. 

A number of points on the blue side of Figure~\ref{jk_tio_mod} seem inconsistent with the 
Phoenix models.  One reason for this may be that the \citet{schlegel98} reddening values
overestimate the reddening, and these points should then come from the most reddened fields.  
We should explore which fields account for these points.  When comparing the \citet{schlegel98}
$\rm E(B-V)$ values for 330 {\sl BRAVA} stars located
within the \citet{gonzalez11} reddening map, a $\sim$0.15 mag offset in $\rm E(B-V)$
is seen, with the \citet{schlegel98} values being systematically larger.  This translates to
$\sim$0.08 mag in $\rm (J-K)$; hence the \citet{gonzalez11} reddening values would
move the $\rm (J-K)_0$ values in Figure~\ref{jk_tio_mod} to the red by roughly the spacing 
between the Phoenix curves, and the number of points on the blue side
of the +0.5 dex Phoenix model would be decreased.  Another possible contribution of 
uncertainties in the TiO strength is the uncertainty in the 2MASS photometry, which
is roughly 0.03 to 0.1 mag in $\rm (J-K)$.  There has also been the long standing problem 
in the bulge \citep{frogel87} that the bulge giants appear to have TiO that is much too 
high for their $\rm (J-K)$.  Further surveys such as the APO Galactic Evolution 
Experiment (APOGEE) will 
obtain high-resolution, high S/N infrared spectroscopy 
of red giants stars across the Galactic bulge, and may shed light on this issue.

\subsection{Metallicity gradients}
\begin{figure}[htb]
\begin{center}
\includegraphics[width=1\hsize]{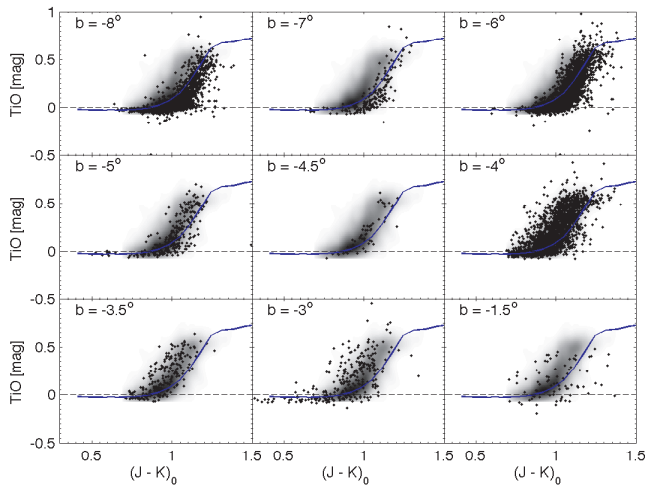}
\end{center}
\caption{Same as Fig.~\ref{jk_tio_mod}, but separated by the Galactic latitudes covered 
by {\sl BRAVA}.  The gray-shaded region indicates the distribution in the $b=-4^{\circ}$ field, 
and the Phoenix model for a metallicity of $-0.5$ dex is over-plotted.  
\label{jk_tio_lat}}
\end{figure}

Recent high resolution spectroscopic studies of the bulge find a vertical abundance gradient in 
the sense of lower metallicities towards higher latitudes \citep[e.g.,][]{zoccali08, johnson11, gonzalez11}. 
Based on the large {\sl BRAVA} sample, this issue can also be investigated (although, here, we 
only do it qualitatively).  Figure~\ref{jk_tio_lat} shows the TiO metallicity indicator 
separated by Galactic latitude.  The Galactic longitude of these fields is $l$=0$^\circ$, and
the Phoenix model for a metallicity of $-0.5$ dex is over-plotted to guide the eye.
It becomes immediately obvious that there is in fact a systematic difference between 
the fields with a tendency for the lower latitude fields to have higher metallicities -- a trend 
consistent with recent suggestions of metallicity gradients across the bulge derived from the K giants
(e.g., Zoccali et al. 2008; Johnson et al. 2011).  
This is also consistent with results of the M giant population from Frogel, Tiede \& Kuchinski 1999;
they use IR CMDs to show there is a gradient in mean metallicity 
along the minor axis in the range 0$^\circ$ $>$ $b$ $>$ $-$10$^\circ$.
Moreover, we find the most prominent trend 
concerns the increase in metallicity spread as one approaches the inner bulge regions.
However, a detailed analysis on the uncertainties in the reddening and photometry in 
the inner bulge regions as compared to those in the outer regions is necessary, especially 
as this is not seen in the high resolution data \citep[i.e.][]{zoccali08, johnson11}.
 
\begin{figure}[htb]
\begin{center}
\includegraphics[width=1\hsize]{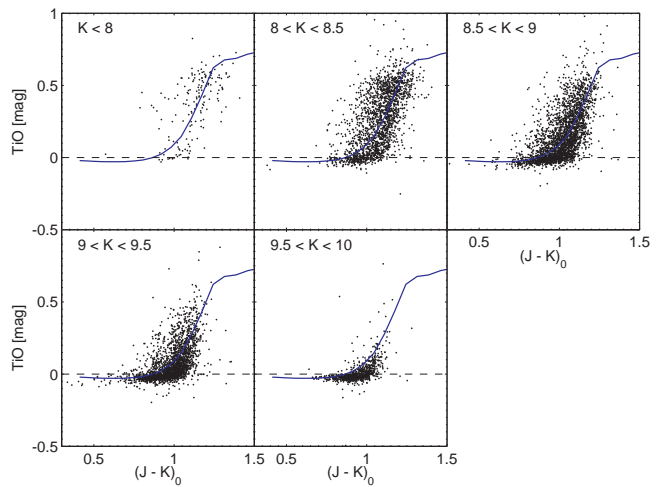}
\end{center}
\caption{Same as Fig.~\ref{jk_tio_mod}, but segregated by stellar magnitude.
\label{jk_tio_mag}}
\end{figure}

Finally, we show in Fig.~\ref{jk_tio_mag} the entire sample segregated by stellar (K-band) magnitude. 
Amongst all of the magnitude bins the color-TiO distributions are broadly consistent with 
each other, bolstering our findings from Section~2.1 that argue against any color/magnitude 
bias in the {\sl BRAVA} sample -- also with regards to metallicity.  The only apparent exception 
is the faintest bin, which, however, exclusively contains stars at the highest latitude (Figure~\ref{cmd}). 
Thus, the metallicity gradient permeates also in this (magnitude-) subsample.

If the \citet{schlegel98} extinction were higher than that of \citet{gonzalez11}, it could in 
principle cause our derived $\rm (J-K)_0$ to be too blue in the fields of greater extinction.
As discussed above, the lower $\rm E(B-V)$ of \citet{gonzalez11} have an 0.08 mag 
effect on $\rm (J-K)_0$.  However, this
is unlikely to mimic a metallicity gradient.  Small-scale variations in reddening are particularly
strong in the inner regions, but these regions 
are mostly constrained to $|b|$$<$4$^{\circ}$ \citep[see][their Fig.~6 and Fig.~7]{gonzalez11}.  
The metallicity gradient in the {\sl BRAVA} data extends all the way
to $b$$=$$-$1.5$^{\circ}$, but is also prominent when considering the $b$$=$$-$8$^{\circ}$ to
$b$$=$$-$4$^{\circ}$ fields.  Further, we find a similar scatter ($\sim$0.1 mag) and a similar
offset ($\sim$0.15 mag) in $\Delta$$\rm E(B-V)$ between
\citet{schlegel98} and \citet{gonzalez11} of our $b$$=$$-$4$^{\circ}$ and $b$$=$$-$6$^{\circ}$
stars.  Yet these fields still show a metallicity gradient.  This strongly suggests that 
reddening uncertainties can not fully explain away the slight metallicity gradient we are seeing.

\section{Outliers}
\begin{figure}[htb]  
\includegraphics[width=1\hsize]{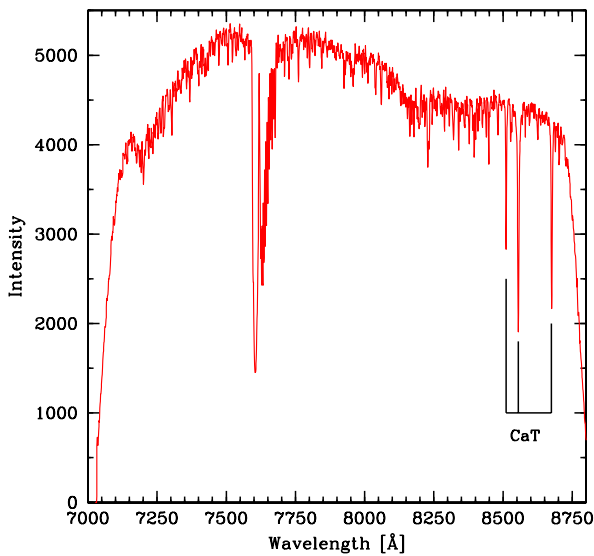}
\caption{Wavelength calibrated spectrum of a velocity outlier in our sample, 
with a Heliocentric velocity of 447 km/s.  The CaT lines are labeled in the
spectrum for reference.
\label{hpv}}
\end{figure}

Radial velocity surveys of the Milky Way halo have the potential to identify rare velocity outliers that 
result from dynamical processes throughout the Galaxy.  
The most obvious outlier in our sample is  2MASS 17464606-3937523, with a heliocentric 
radial velocity of 447 km/s.  Its spectrum is shown in Figure~\ref{hpv}.  
The S/N is high, and to first order it looks metal poor.  Unfortunately, at 
($l$,$b$)=($-$9.0433,$-$5.7386), it is not in the OGLE-II proper motion catalog \citep{sumi04}.  
Hypervelocity stars (HPVs) were first discovered by \citet{brown05}, and are generally
B-type stars moving 2-3 times the Galactic escape velocity.  At a heliocentric velocity of 
$\sim$450 km/s, the star may not qualify as a true hypervelocity star, although at $r$=50 to 100 kpc, 
unbound stars have such velocities \citep{kenyon08}.  A preliminary $\rm [Fe/H]$ based on the CaT 
is $-$0.86$\pm$0.05 dex, with an error purely based on the continuum variance
and magnitude errors, not accounting for uncertainties in the calibration coefficients.  This
is not an unusual $\rm [Fe/H]$ value for a bulge star.

\section{Results and Analysis}
\begin{figure}[htb]  
\includegraphics[width=1\hsize]{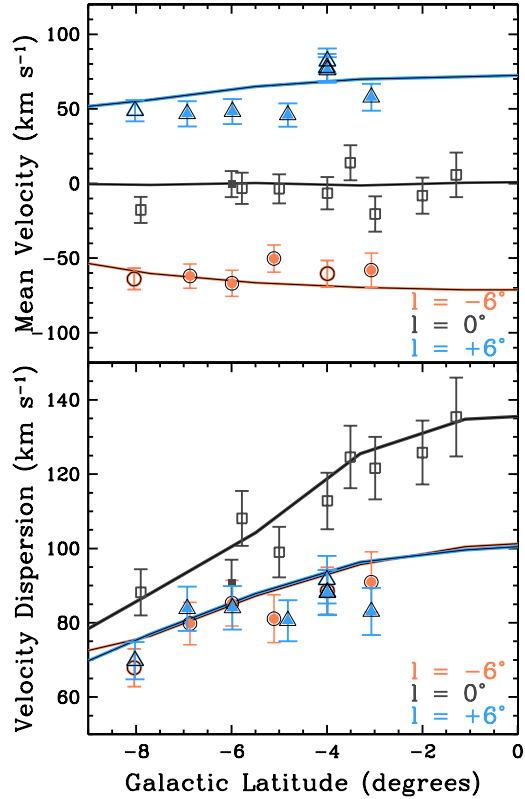}
\caption{Velocity dispersion profile (bottom) and rotation curve (top) for the $l$=$-$6$^\circ$,
+6$^\circ$ and 0$^\circ$ strips.  The open symbols indicate data already published and the 
filled symbols indicate the data presented here.  The model of \citet{shen10} is overplotted, 
with good agreement to the observations.
\label{lm6strip1}}
\end{figure}

Radial velocities, especially in minor axis fields, can constrain bulge models. 
Figure~\ref{lm6strip1} shows the dispersion profile and rotation curve for the minor-axis strips
at $l=-6^\circ$, $l=6^\circ$, and $l=0^\circ$.  The predictions of the \citet{shen10} models is 
over-plotted.
The Shen model uses a cylindrical particle-mesh code to construct an $N$-body model to the
{\sl BRAVA} data. The initial parameters adopted are an unbarred disk and a rigid pseudo-isothermal
halo potential \citep[see][for details]{shen10}.  In this model, the Milky Way self-consistently develops
a bar, which buckles and thickens in the vertical directions.  Hence, there is no classical bulge
component, and the best-fit model predicts a bar half-length of $\sim$4 kpc, extending 20$^\circ$ from
the Sun-Galactic center line.   The additional data presented here allows the Shen model to be 
compared to 
two more minor axis fields.  As seen in \citet{shen10}, the data agree well with the Shen model 
for both the mean velocity and velocity dispersion observations.  

\begin{figure}[htb]  
\includegraphics[width=1\hsize]{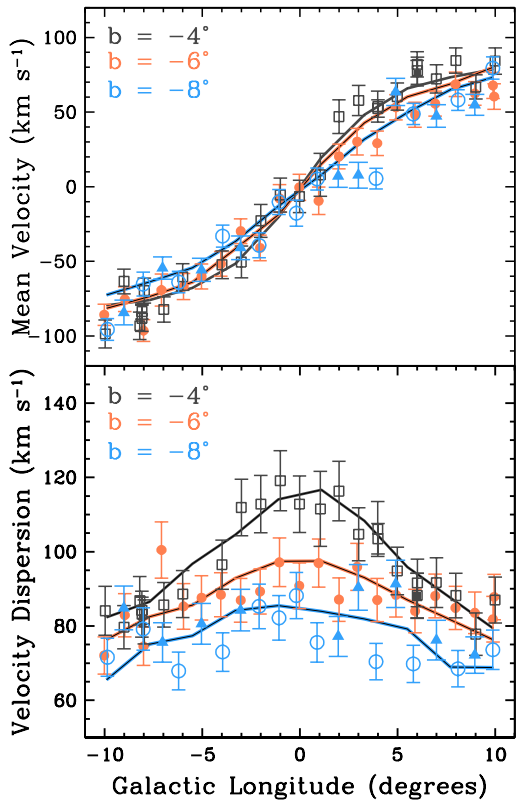}
\caption{Velocity dispersion profile (bottom) and rotation curve (top) for the $b$=$-$4$^\circ$,
$-$6$^\circ$ and $-$8$^\circ$ strips.  The filled symbols indicate data already published and 
the open symbols indicate the data presented here.
\label{bm6strip1}}
\end{figure}

Figure~\ref{bm6strip1} shows the dispersion profile and rotation curve for the major-axis strips
at $b=-4^\circ$, $b=-6^\circ$, and $b=-8^\circ$.  The $b=-4^\circ$ data comes from Paper I 
nd half of the $b=-8^\circ$
data comes from \citet{howard09}.   The Shen model matches the observations well, validating
the results of \citet{shen10}.  It is also worth noting that the rotation curves at $b=-4^\circ$, 
$-6^\circ$, and $-8^\circ$ are 
remarkably similar, suggesting that the Galaxy's bulge rotates cylindrically.   
This had been suggested by \citet{howard09}, but here the observational data are more 
plentiful and include additional fields at $b=-$6$^\circ$.
Thus our bolstering of the cylindrical rotation signal is further evidence that the bulge is an 
edge-on bar, as predicted by the Shen model.  

\begin{figure}[htb]  
\includegraphics[width=1\hsize]{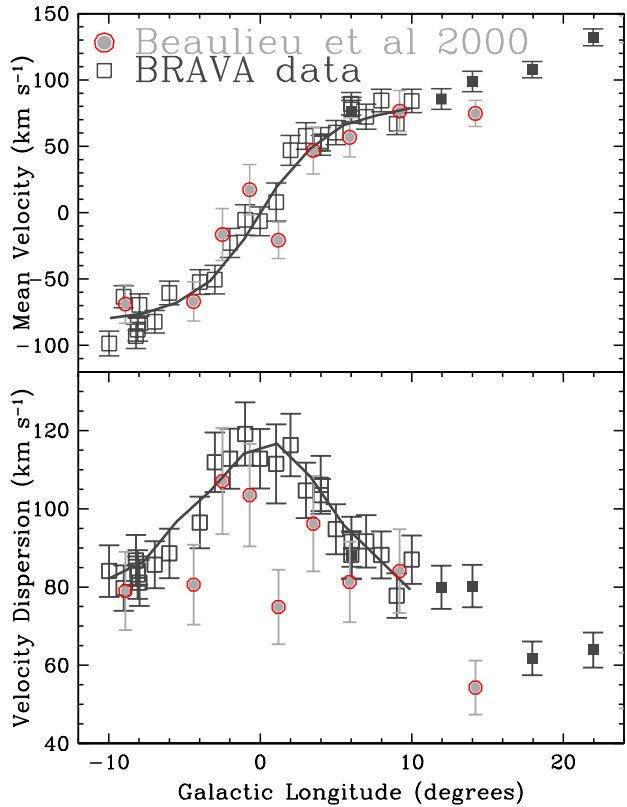}
\caption{Velocity dispersion profile (bottom) and rotation curve (top) for all the $b$=$-$4$^\circ$
fields.  The filled symbols indicate data already published and the open symbols indicate
the data presented here.  Over-plotted is data from bulge PNes from \citet{beaulieu00}, as
well as the predictions of the \citet{shen10} model.
\label{disk}}
\end{figure}
The velocity dispersion drops as one moves further in latitude from the Galactic disk plane.
A hotter model contributes to a large velocity dispersion, and the nucleus-spheroid component 
of the model is hot.  However, the {\sl BRAVA} data show remarkable agreement with the disk/bar 
component of the model, with a relatively flat dispersion profile at $\sim$70 km s$^{-1}$ 
contrasting with the spheroid dispersion, which is predicted to be at $\sim$120 km s$^{-1}$. 
The turnover seen in the rotation curve at $b$ = $-$4$^\circ$ is not evident at $b$ = $-$6$^\circ$
or $b$ = $-$8$^\circ$. 

Additionally, in Figure~\ref{disk}, the {\sl BRAVA} fields extending past the main 
body of the bulge at $l$$>$10$^\circ$ are plotted.  These observations lie outside the high
surface brightness boundary of the COBE bulge, and can be used to examine the
extent of the bar/bulge and to look for signs of disk contamination.  
The full length of the bar is thought to be on the order of 
3-4 kpc \citep{hammersley00, bissantz02, cabrera07}, 
and since the bar's pattern speed is relatively 
rapid, it effectively ``controls" the region that it lives in.  
At the distance of the Galactic Center (7.9 kpc), 4 kpc at the bar angle of $\sim$20 degrees
corresponds to $\pm l$$\sim$16$^{\circ}$.  Indeed, the fields at $l$$>$15$^\circ$ show a drop
in velocity dispersion as well as a higher mean velocity, which is due presumably
to the presence of the inner disk component in these fields.   Figure~\ref{disk}
also shows the observations from a sample of 373 PNes \citep{beaulieu00};
the {\sl BRAVA} data confirms
the rotation seen by the PNes at $|l|$ $>$ 12$^{\circ}$,
as well as a drop in the velocity dispersion in these fields.  The region covered
in the \citet{beaulieu00} data is $-$20$^{\circ}$ $<$ $l$ $<$ 20$^{\circ}$ and
$-$5$^{\circ}$ $>$ $b$ $>$ $-$10$^{\circ}$. Observations
at $|l|$ $>$ 12$^{\circ}$ place
important constraints for Galactic bulge theoretical models and predictions, such as the bar angle 
(Martinez-Valpuesta, private communication).

\begin{figure}[htb]
\includegraphics[width=1\hsize]{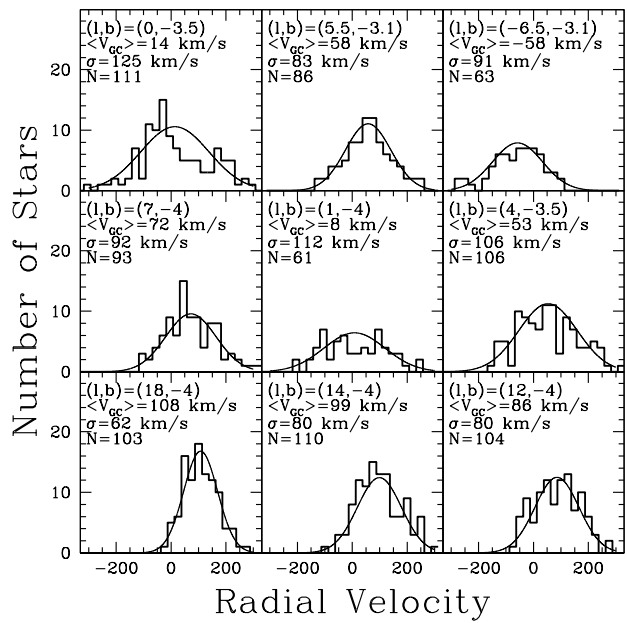}
\caption{Presentation of all bulge field galactocentric velocity distributions. Overlaid on each 
plot is a Gaussian derived from the field statistics.
\label{RVf}}
\figurenum{8}
\end{figure}

\begin{figure}[htb]  
\includegraphics[width=1\hsize]{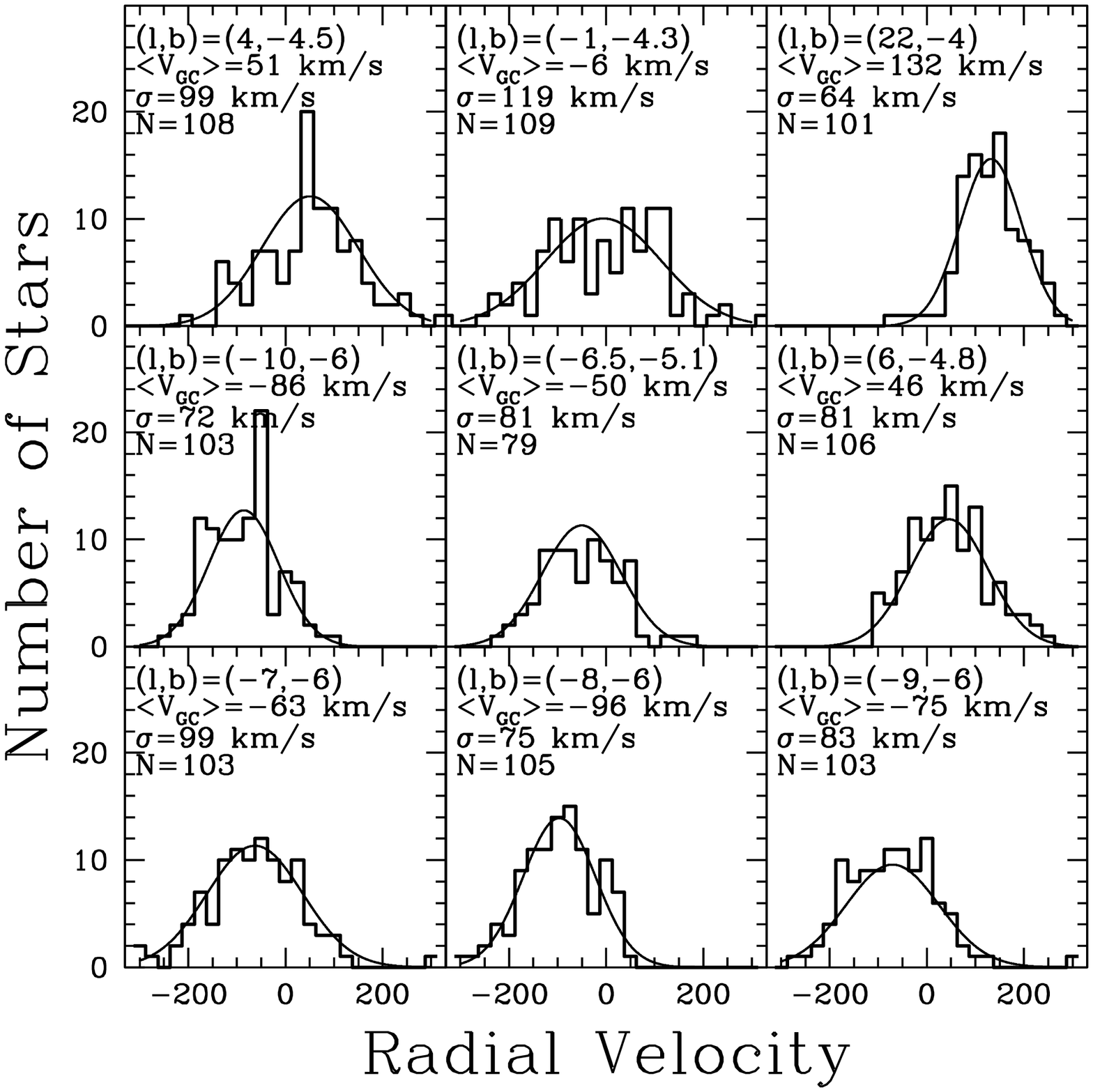}
\caption{Bulge line-of-sight velocity distributions continued
\label{RVe}}
\figurenum{8}
\end{figure}

\begin{figure}[htb]  
\includegraphics[width=1\hsize]{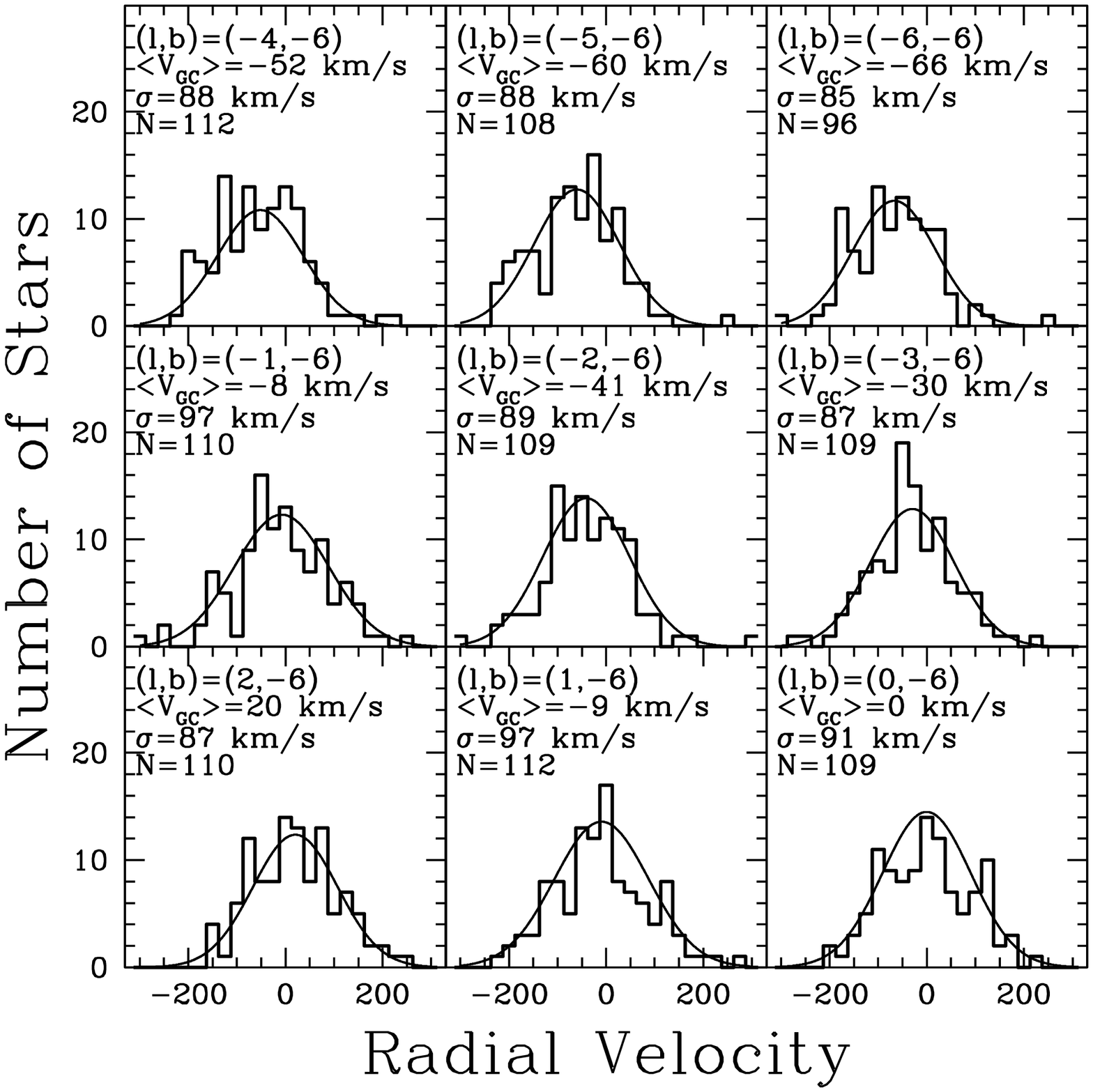}
\caption{Bulge line-of-sight velocity distributions continued
\label{RVd}}
\figurenum{8}
\end{figure}

\begin{figure}[htb]  
\includegraphics[width=1\hsize]{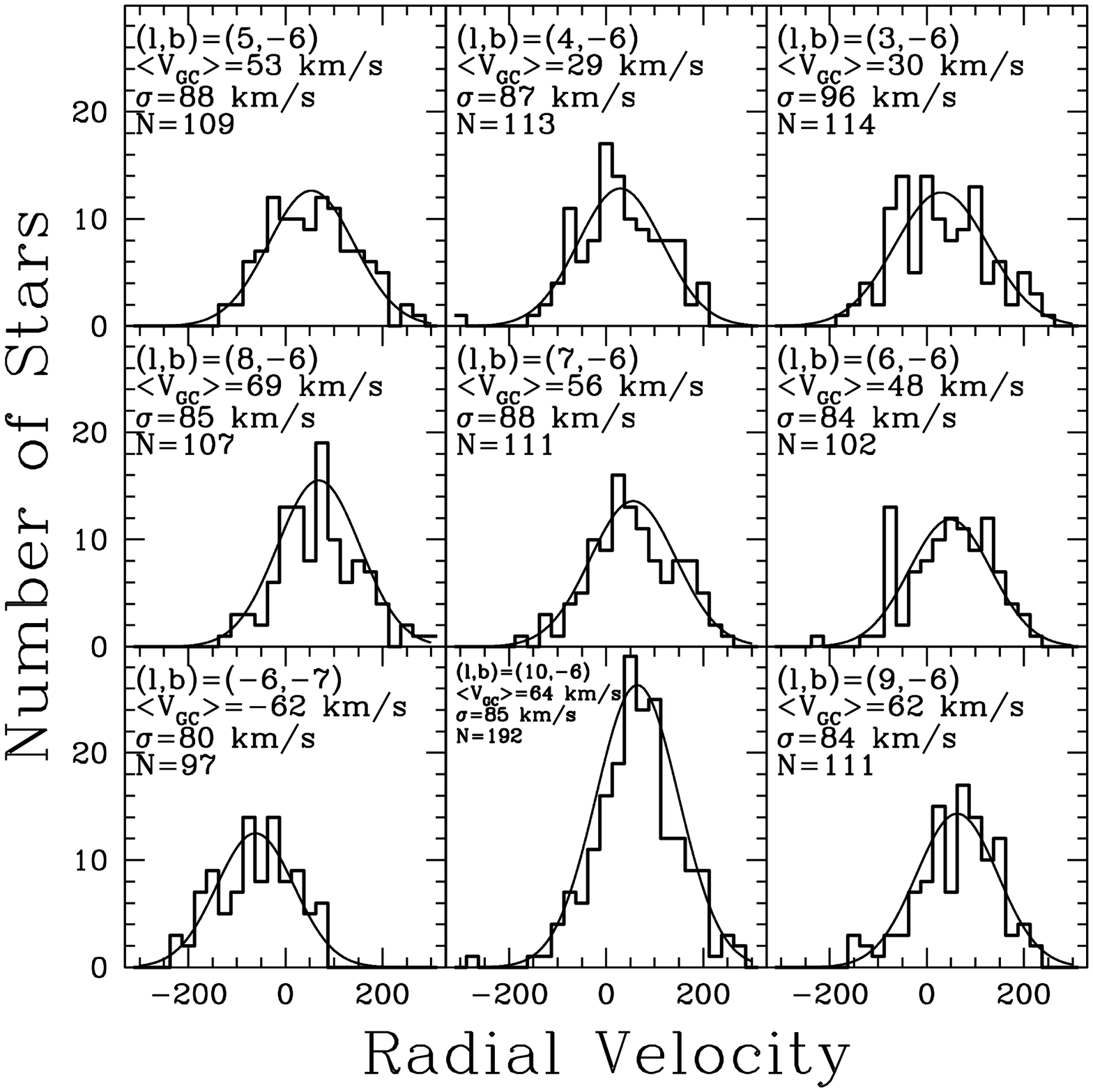}
\caption{Bulge line-of-sight velocity distributions continued
\label{RVc}}
\figurenum{8}
\end{figure}

\begin{figure}[htb]  
\includegraphics[width=1\hsize]{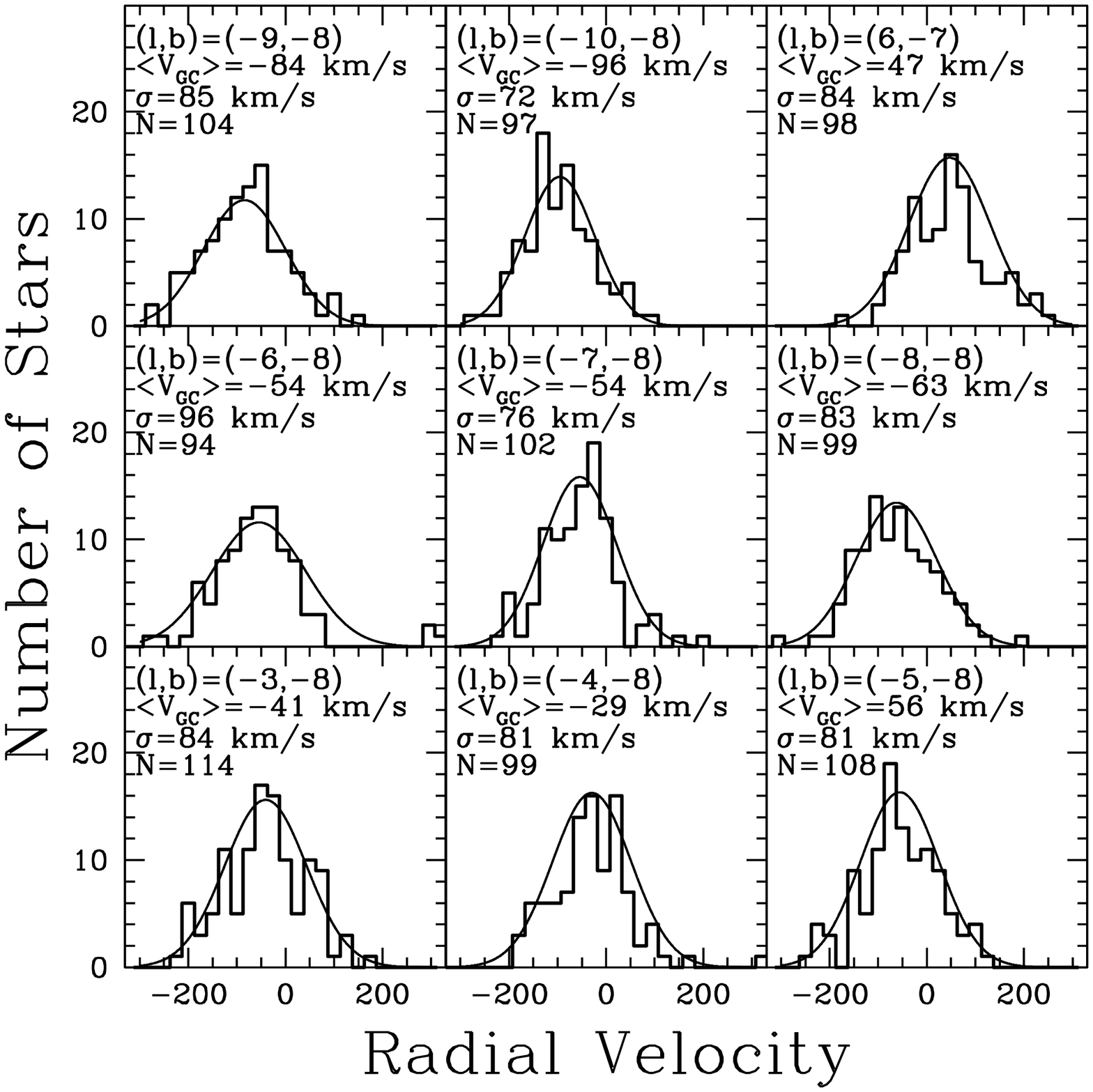}
\caption{Bulge line-of-sight velocity distributions continued
\label{RVb}}
\figurenum{8}
\end{figure}

\begin{figure}[htb]  
\includegraphics[width=1\hsize]{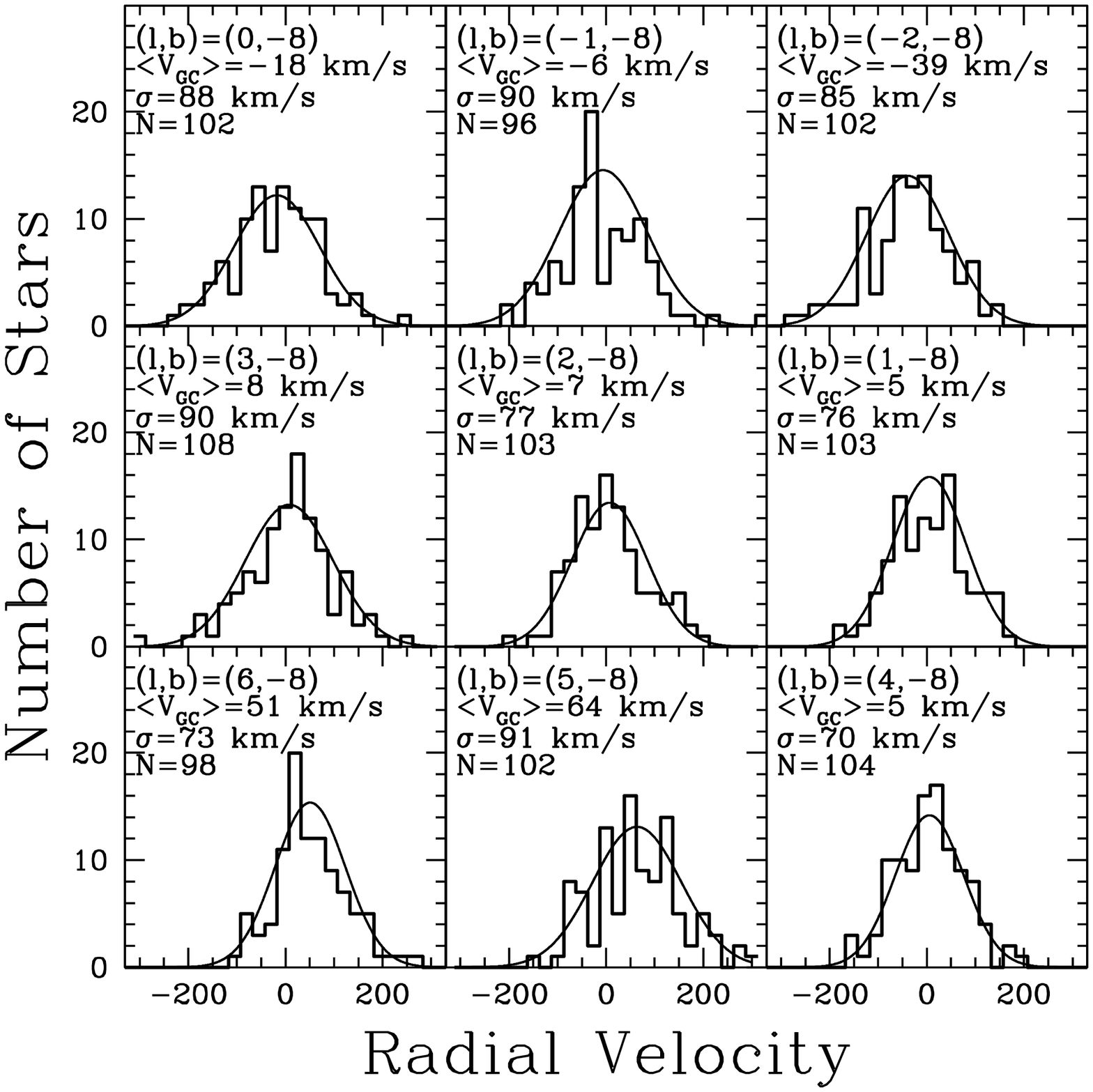}
\caption{Bulge line-of-sight velocity distributions continued
\label{RVa}}
\figurenum{8}
\end{figure}

\begin{figure}[htb]  
\includegraphics[width=1\hsize]{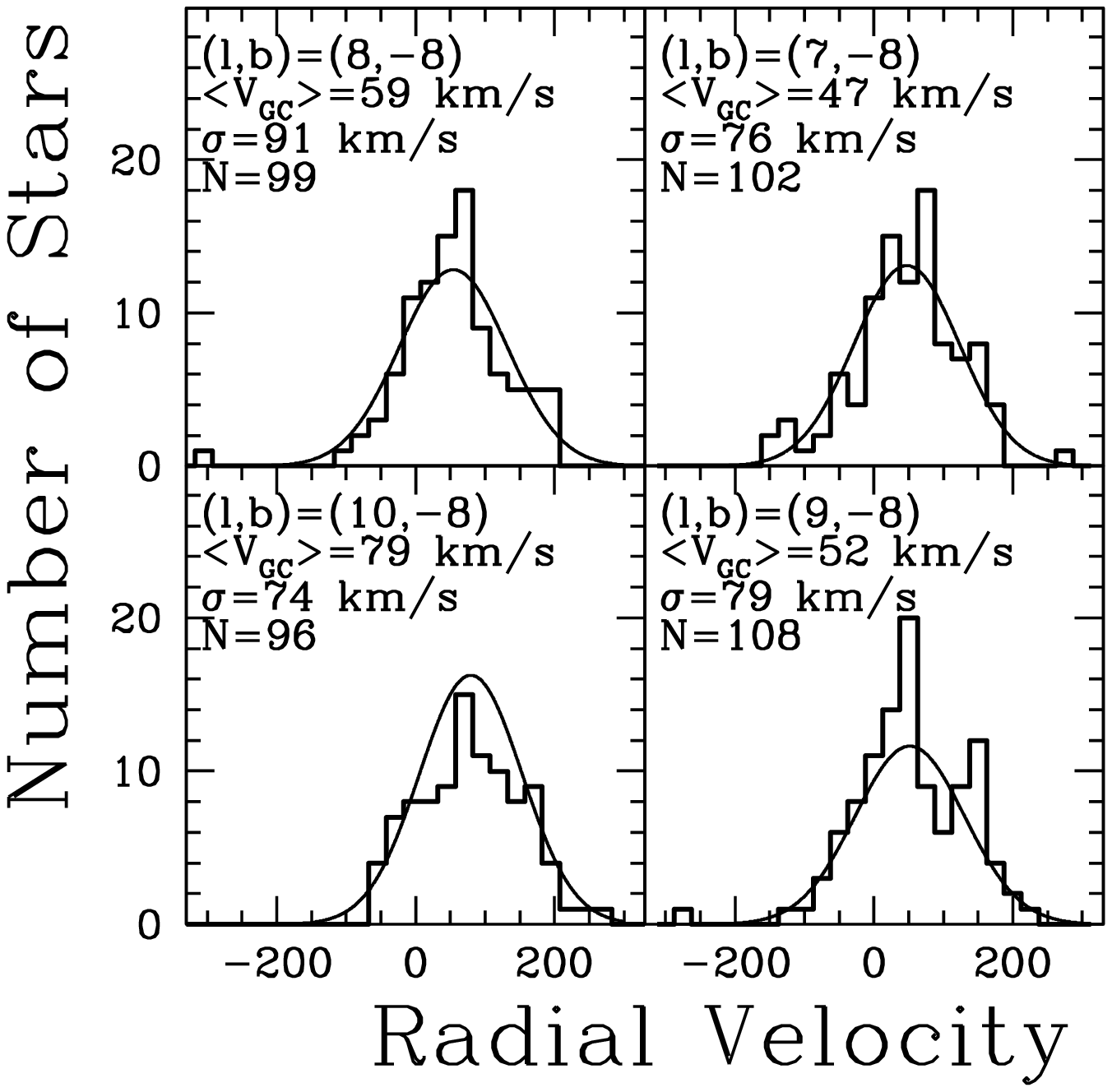}
\caption{Bulge line-of-sight velocity distributions continued
\label{RVaa}}
\figurenum{8}
\end{figure}

The galactocentric velocity distributions of the new bulge fields are now presented.
As with Paper I, a bin width of 25 km s$^{-1}$ is used, driven by the approximate dispersion 
expected for cold components like a dissolving cluster or dwarf galaxy ($\sim$10 km/s).  
Simulations by \citet{reitzel07} show that with the number of stars we have in each 
field ($\sim$100), deviations from a normal distribution are not unexpected in random draws.  
However, fields with the largest deviations from a normal distribution may aid in the selection 
of follow-up observations.  To test for normality, the Shapiro-Wilk 
test (Shapiro \& Wilk 1965; D'Agostino 1986) and
the Anderson-Darling test \citep{press86} are employed.  These normality tests are 
two of the most powerful tests for deviations from normality and are especially useful for
sample sizes that are relatively small, i.e., $\sim$100 particles.  For example, 
the Shapiro-Wilk normality test
has successfully been used by \citet{kepley07} to detect known streams 
in one component of velocity.  
For both of these tests, the smaller the $p$-value, the less likely it is that the data come from 
a normal distribution, and a $p$-value $<$ 0.05 rejects the supposition of normality.
The results of these tests are given in Table~\ref{normality}.
The tested {\sl BRAVA} field is first listed in Table~\ref{normality} followed by the number
of stars in the field, the skewness and kurtosis, and its associated 
$p$-value from the Shapiro-Wilk test and from the Anderson-Darling test.  

In general, the skewness and kurtosis values are all within one or two sigma, the exception
being fields ($l$,$b$) =  ($-$1,$-$8), ($-$4,$-$8), ($-$6,$-$8), (8,$-$8) and (9,$-$8). 
Upon closer examination, it is apparent that except for  ($l$,$b$) =  ($-$6,$-$8) 
which is discussed in more detail below, these fields each contain a star on the tail end of the 
distribution which greatly influences the skewness and kurtosis values.  
These fields also all 
fail the Shapiro-Wilk normality test, which gives more weight to the tails of the distribution
than the Anderson-Darling test.  All of these fields
are located at $b$=$-$8$^\circ$, and it is likely that we are seeing some signatures
of the bulge/disk boundary in these fields.

There are only two {\sl BRAVA} fields that fail both the Shapiro-Wilk and the 
Anderson-Darling test.  
These fields are located at ($l$,$b$) =  (0,$-$3.5) and ($l$,$b$) =  ($-$6,$-$8).
The former has a skewness and kurtosis consistent with a Gaussian distribution, but
$p$-values that reject the supposition of normality.  Removing the stars that lie
in the tails of the distribution does not cause the $p$-value to increase and 
hence indicate normality.  There is a strong peak
in the velocity distribution at around $-$30 km/s and also not many stars at $\sim$100 km/s,
which are the likely features causing the
Shapiro-Wilk and Anderson-Darling tests to reject normality.
Interestingly, the possible signature of a disrupted satellite reported by \citet{rangwala09}
at ($l$,$b$) =  (+5.5,$-$3.5), has a velocity distribution excess around $-$35 km s$^{-1}$.
However, we find no correlation at ($l$,$b$) =  (0,$-$3.5) with the stars that have 
velocities around $-$30 km/s and
their TiO values.  Our sample size is small, and follow-up observations of this field
would be particularly interesting.

The field at ($l$,$b$) =  ($-$6,$-$8), on the other hand, has the largest skewness and kurtosis
of any {\sl BRAVA} fields (1.38 $\pm$ 0.25 and 4.91 $\pm$ 0.51, respectively).  
Further, it has three stars with heliocentric velocities, $\rm V_{HC}$, greater than 300
km s$^{-1}$; the radial velocity dispersion of these three stars is 16 $\pm$ 7 km/s.
From our complete sample of 8585 stars, only 11 stars have velocities that are
above 300 km/s, and so it is striking that three of these stars are located in this field. 
Again it would be especially interesting to obtain follow-up observations of the stars in this field 
to see if more stars with such large velocities are found.

Figure~\ref{RVall} shows all the 8585 radial velocities obtained from the {\sl BRAVA} dataset 
co-added (bottom panel).  The mean is 2 $\pm$ 1 km s$^{-1}$ with a 
$\sigma$ $=$ 107 $\pm$ 1 km s$^{-1}$.  
The skewness is negligible (0.03 $\pm$ 0.03) and the kurtosis of  $-$0.17 $\pm$ 0.05 implies 
a slightly platykurtic (flattened) distribution.  Both of these results are in agreement with the 
results found by Paper I.  Further, the lack of significant skewness and a small value of kurtosis
is consistent with our argument that our Bulge radial velocity distribution is not largely
contaminated by either cold components (disk) or hot components (halo).  

A Shapiro-Wilk
normality test fails for this co-added dataset, as the sample is too large; with 8585 stars, there
are many radial velocities with very similar values.  An Anderson-Darling normality test
gives a $p$-value of 0.59 and a Kolmogorov-Smirnov test, which works best with large
datasets, gives a $p$-value of 0.10.  
These results suggest no significant deviations from a Gaussian distribution.

The top panel of Figure~\ref{RVall} shows the radial velocities shifted
onto zero and then co-added.  The mean is $-$3 $\pm$ 1 km s$^{-1}$ with 
a $\sigma$ $=$ 95 $\pm$ 1 km s$^{-1}$.  
Again, the distribution is Gaussian, with no apparent deviation from a normal distribution.
It is noteworthy that there is only one star in our sample with a velocity greater 
than $\pm$4 $\sigma$, which is discussed in \S4.  

Figure~\ref{l_v} shows the longitude-velocity plot for the three {\sl BRAVA} major axis strips.
For latitudes closest to the plane ($b$=$-$4$^\circ$ 
and  $b$=$-$6$^\circ$) there is no evidence of a cold, disk component 
in our sample which would manifest itself as a linear trend.  However, further from the 
plane ($b$=$-$8$^\circ$, $\sim$1.2 kpc from the Galactic
plane), the ``S" shape is not as apparent, suggesting that the disk
does contribute to the {\it BRAVA} sample in this regime.  Indeed as compared
to the $b$=$-$4$^\circ$ and $b$=$-$6$^\circ$ fields, the $b$=$-$8 $^\circ$
fields are also those that in general have skewness, kurtosis and
$p$-values that are least consistent with a normal distribution (see Table~\ref{normality}).
Recent studies of planetary nebulae in the inner MW suggest a bulge-disk interface
at $\sim$1.5 kpc \citep{cavichia11}, consistent with our findings.

Figure~\ref{b_v} shows the latitude-velocity plot for three {\sl BRAVA} minor axis strips.
As expected a linear trend is seen; neither cold components
nor indications of a hot, non-rotating population is seen.  However we see "rotation" in the
sense of solid body rotation.
\setcounter{figure}{13} 
\begin{figure}[htb]  
\includegraphics[width=1\hsize]{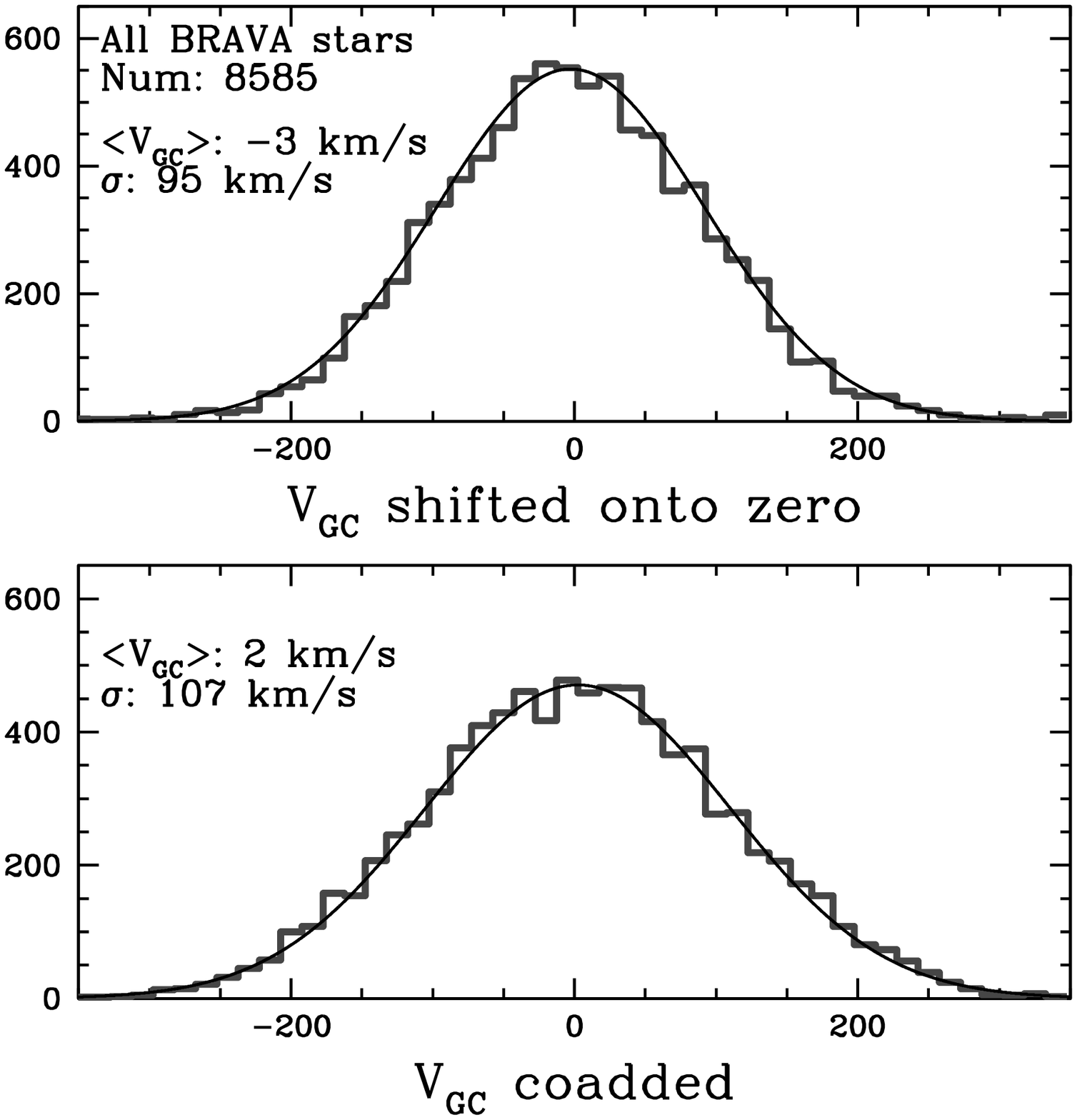}
\caption{The co-added galactocentric radial velocities of the {\sl BRAVA} stars.  In the top
panel, each {\sl BRAVA} field is shifted to zero before co-addition.  A bin width of 
15 km s$^{-1}$ is used for both histograms.  Neither distributions show any signs of
deviation from a Gaussian.
\label{RVall}}
\end{figure}

\setcounter{figure}{14} 
\begin{figure}[htb]  
\includegraphics[width=1\hsize]{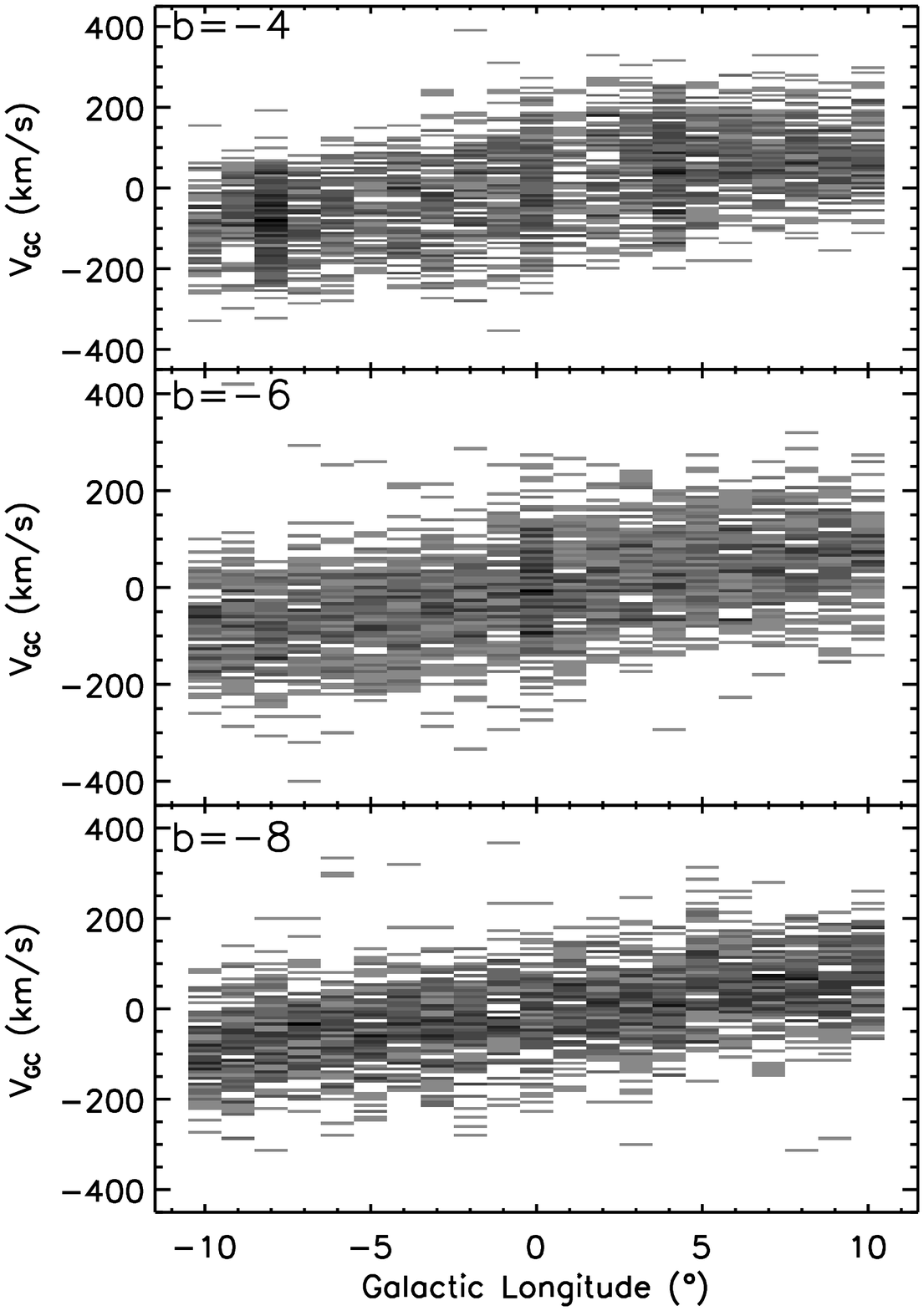}
\caption{Longitude-Velocity (\rm $l_v$) plot for the entire bulge sample at 
$b$=$-$4$^\circ$ (top), $b$=$-$6$^\circ$ (middle) and $b$=$-$8$^\circ$
(bottom).  The $l_v$ plot is smoothed to 1$^\circ$ in longitude and 
10 km s$^-1$ in galactocentric velocity.  This figure shows the cylindrical rotation 
trend very clearly.  Notice the lack of any prominent ``cold" features that would indicate a 
possible stream detection across multiple fields.
\label{l_v}}
\end{figure}

\setcounter{figure}{15} 
\begin{figure}[htb]  
\includegraphics[width=1\hsize]{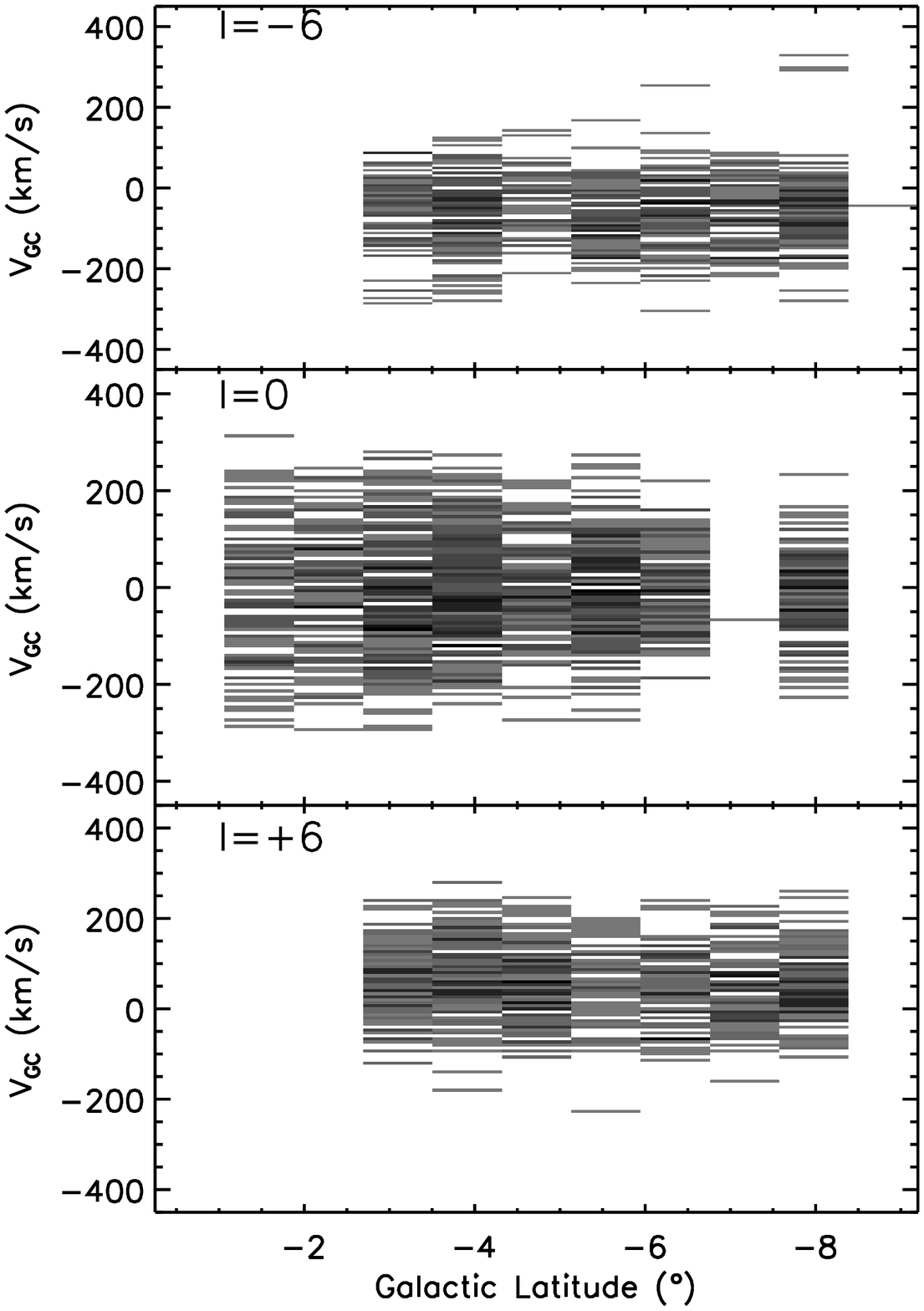}
\caption{Latitude-Velocity (\rm $b_v$) plot for the entire bulge sample at 
$l$=$-$6$^\circ$ (top), $l$=0$^\circ$ (middle) and $b$=$+$6$^\circ$
(bottom).  The sample is smoothed to 1$^\circ$ in latitude.
\label{b_v}}
\end{figure}

\subsection{A Milky Way Pseudobulge?}
A vertical metallicity gradient in the bulge has been reported in
several recent studies \citep{melendez08, zoccali08, babusiaux10, johnson11}, 
and it has often been
used as an argument against the boxy bulge/bar model of \citet{shen10}. 
It is true that the collective bar buckling happens to
essentially the whole disk that is in place at the time of the
buckling instability, but a vertical metallicity gradient could still be
consistent with the Shen bar/pseudobulge model. One possibility is
that some of the vertical thickening could be produced by resonant
heating of stars that scatter off the bar \citep{pfenniger84, pfenniger85, pfenniger90}.
If the most metal-poor stars are also the oldest stars, then they have
been scattered for the longest time and now reach the greatest
heights, hence giving rise to a vertical metallicity gradient.
Secondly, the \citet{shen10} model still allows a relatively
small merger-built classical bulge (about 10\% of the disk mass), the
mixing of two bulge populations could conceivably produce a vertical
metallicity gradient (also see the Besancon model, Robin et al. 2011).
A mixing of the bar/pseudobulge with the inner thick disk was also
proposed to explain the vertical metalicity gradient \citep{bekki11}.

That the Milky Way has a pseudobulge is consistent with
the majority of the bright galaxies in the local 11 Mpc volume 
\citep{kormendy10n, fisher11}, especially
for those galaxies with a mass similar to that of the Milky Way.
Recently an in-depth study of NGC$\,$4565 has revealed that this galaxy also contains a 
pseudobulge and no hint of a merger-built classical bulge \citep{kormendy10}.
Like the MW, NGC$\,$4565 has a peanut shaped bulge which rotates cylindrically 
\citep{kormendy82} and has a strong abundance gradient \citep{proctor00}.  Hence 
other galaxies besides the MW
with an abundance gradient but no (or very little) classical bulge are known to exist.
Evolution of the vertical metallicity gradient however has not been extensively 
investigated by theoretical studies, but the model by \citet{bekki11} is a step forward
in understanding why an abundance gradient is observed in the bulge even if the bar
evolved secularly.

\section{Conclusions}
We have presented the final data set of the {\sl BRAVA} survey and provided a website for data 
access, which is available at the IRSA archive:
{\tt http://irsa.ipac.caltech.edu/} as well as at UCLA: {\tt http://brava.astro.ucla.edu/}.
The observations at the $b$=$-$6$^{\circ }$ latitude fields as well as more observations
at the $b$=$-$8$^{\circ }$ latitude fields confirms previous suggestions that the Galactic bulge
rotates cylindrically, as do boxy bulges of other galaxies \citep[e.g.,][]{kormendy82, 
jarvis90, shaw93, kormendy04, howard09}.  
The data in this paper double the number of stars shown in \cite{howard08,howard09} and used
in \cite{shen10}.  Analysis of these data in the same fashion shows that the Galactic bulge consists
of a single massive bar formed by secular evolution. The existence of such a massive bar and
no classical bulge implies that our Galaxy has not suffered any large mergers (enough to form
a classical bulge) since the epoch at which the disk formed. 
Based on TiO$\varepsilon$ index measurements we qualitatively see the signatures 
of a vertical metallicity gradient -- an attribute that is in contrast to our kinematic  
characterization of the bulge as a pure pseudobulge. A detailed analysis of the {\sl BRAVA} 
metallicity scale and its spatial distribution is clearly warranted and underway. 

\acknowledgments
It is our pleasure to thank Inma Martinez-Valpuesta and Ortwin Gerhard for
helpful discussions.  Nine of us (A. Kunder, R.M. Rich, C.I. Johnson, J. Shen, A.C. Robin, 
M. Soto, P. Frinchaboy, Y. Wang and L. Origlia) wish to acknowledge the hospitality of the 
Aspen Center for Physics, which is supported by the NSF grant \#1066293, and where
much dialogue and exchange of ideas took place.  A. Kunder thanks Dan Phillips for his 
work on the {\sl BRAVA} website.
A. Koch thanks the Deutsche Forschungsgemeinschaft for funding from Emmy-Noether grant  
Ko 4161/1.  
This material is based upon work supported by the National Science Foundation under
awards AST--0709479 to R. M. Rich and No. AST--1003201 to C.I.J.
The research presented here is
partially supported by the National Natural Science Foundation of
China under grant No. 11073037 to JS and by 973 Program of China under
grant No. 2009CB824800 to JS.  We thank the anonymous referee for suggestions
that strengthened our analysis.

\clearpage

\begin{table}
\begin{scriptsize}
\centering
\caption{{\sl BRAVA} Rotation and Dispersion Results}
\label{obs}
\begin{tabular}{p{0.67in}p{0.5in}p{0.55in}p{0.24in}p{0.24in}p{0.45in}p{0.5in}p{0.5in}p{0.5in}p{0.5in}p{0.4in}p{0.4in}} \\ \hline
Date of Obs (UT) & R.A. (J2000.0) & Decl. (J2000.0) & gal $l$ (deg) & gal $b$ (deg) & Target Fibers in Config & Reliable Velocities & $\rm <V_{HC}>$ (km s$^{-1}$) & $\rm <V_{GC}>$ (km s$^{-1}$) & err$_{<V>}$ (km s$^{-1}$) & $\sigma$ (km s$^{-1}$) & err$_{\sigma}$ (km s$^{-1}$) \\ 
\hline
2008 Jul 08 & 17:57:05.75 & -40:43:23.4 & $-$9.0 & $-$8.0 & 108 & 104 & $-$56.10 & $-$84.26 & 8.32 & 84.86 & 5.88 \\
2008 Jul 08 & 18:02:09.79 & -38:59:18.5 & $-$7.0 & $-$8.0 & 102 & 102 & $-$34.14 & $-$54.39 & 7.49 & 75.60 & 5.29 \\
2008 Jul 08 & 18:14:40.35 & -25:42:48.7 & \ \ 6.0 & $-$4.0 & 107 & 106 & \ \ 43.38 & \ \ 76.03 & 8.56 & 88.12 & 6.05 \\
2008 Jul 08 & 18:36:28.77 & -24:54:15.4 & \ \ 9.0 & $-$8.0 & 109 & 108 & \ \ 11.27 & \ \ 54.85 & 6.95 & 72.20 & 4.91 \\
2008 Jul 09 & 18:06:51.74 & -37:15:44.5 & $-$5.0 & $-$8.0 & 110 & 108 & $-$43.43 & $-$55.69 & 7.76 & 80.62 & 5.49 \\
2008 Jul 09 & 18:11:25.87 & -35:30:15.1 & $-$3.0 & $-$8.0 & 115 & 114 & $-$36.62 & $-$40.84 & 7.89 & 84.27 & 5.58 \\
2008 Jul 09 & 18:22:18.76 & -31:05:47.0 & \ \ 2.0 & $-$8.0 & 105 & 103 & \ $-$8.65 & \ \ \ \ 7.16 & 7.61 & 77.21 & 5.38 \\
2008 Jul 09 & 18:24:32.95 & -30:12:46.9 & \ \ 3.0 & $-$8.0 & 109 & 108 & $-$12.17 & \ \ \ \ 7.74 & 8.70 & 90.39 & 6.15 \\
2008 Jul 09 & 18:28:19.59 & -28:29:36.9 & \ \ 5.0 & $-$8.0 & 106 & 102 & \ \ 35.93 & \ \ 63.48 & 9.04 & 91.34 & 6.39 \\
2008 Jul 10 & 17:41:55.43 & -36:03:35.9 & $-$6.5 & $-$3.1 & 65 & 63 & $-$40.86 & $-$58.03 & 11.46 & 90.96 & 8.10 \\
2008 Jul 10 & 17:50:54.07 & -37:02:24.9 & $-$6.5 & $-$5.1 & 89 & 79 & $-$33.10 & $-$50.23 & 9.12 & 81.09 & 6.45 \\
2008 Jul 10 & 18:10:07.40 & -25:39:30.9 & \ \ 5.5 & $-$3.1 & 88 & 86 & \ \ 26.54 & \ \ 57.80 & 8.95 & 83.02 & 6.33 \\
2008 Jul 10 & 18:32:36.91 & -26:39:15.7 & \ \ 7.0 & $-$8.0 & 104 & 102 & \ \ 11.58 & \ \ 47.41 & 7.55 & 76.21 & 5.34 \\
2008 Jul 11 & 17:59:22.08 & -37:39:25.0 & $-$6.0 & $-$6.9 & 102 & 97 & $-$45.49 & $-$61.92 & 8.10 & 79.79 & 5.73 \\
2008 Jul 11 & 18:17:49.98 & -26:10:14.0 & \ \ 6.0 & $-$4.8 & 107 & 106 & \ \ 13.82 & \ \ 45.92 & 7.83 & 80.56 & 5.53 \\
2008 Jul 11 & 18:22:15.60 & -26:41:40.5 & \ \ 6.0 & $-$6.0 & 104 & 102 & \ \ 16.35 & \ \ 48.22 & 8.32 & 84.03 & 5.88 \\
2008 Jul 11 & 18:26:03.24 & -27:07:50.2 & \ \ 6.0 & $-$6.9 & 99 & 98 & \ \ 14.94 & \ \ 46.70 & 8.46 & 83.78 & 5.98 \\
2008 Jul 11 & 18:29:57.01 & -23:06:19.2 & \ 10.0 & $-$5.9 & 93 & 88 & \ \ 20.85 & \ \ 68.56 & 8.72 & 81.54 & 6.17 \\
2008 Jul 12 & 17:55:51.60 & -37:05:52.7 & $-$6.0 & $-$6.0 & 102 & 96 & $-$51.06 & $-$66.83 & 8.70 & 85.28 & 6.15 \\
2008 Aug 18 & 17:45:20.95 & -40:34:09.7 & $-$10.0 & $-$6.0 & 104 & 103 & $-$53.70 & $-$85.72 & 7.10 & 72.09 & 5.02 \\
2008 Aug 19 & 17:48:04.97 & -39:41:08.2 & $-$9.0 & $-$6.0 & 105 & 103$^{a}$ & $-$47.46 & $-$75.29 & 8.21 & 82.91 & 5.81 \\
2008 Aug 19 & 17:50:44.04 & -38:51:51.5 & $-$8.0 & $-$6.0 & 106 & 105 & $-$72.36 & $-$96.35 & 7.27 & 74.54 & 5.14 \\
2008 Aug 19 & 17:53:05.49 & -38:00:47.4 & $-$7.0 & $-$6.0 & 111 & 103 & $-$42.71 & $-$62.86 & 9.71 & 98.50 & 6.86 \\
2008 Aug 19 & 18:26:34.99 & -24:52:48.8 & \ \ 8.0 & $-$6.0 & 109 & 107 & \ \ 28.44 & \ \ 68.56 & 8.21 & 84.88 & 5.80 \\
2008 Aug 20 & 17:58:08.76 & -36:17:31.2 & $-$5.0 & $-$6.0 & 109 & 108 & $-$48.01 & $-$60.06 & 8.43 & 87.60 & 5.96 \\
2008 Aug 20 & 18:28:33.61 & -23:59:55.1 & \ \ 9.0 & $-$6.0 & 111 & 111 & \ \ 18.41 & \ \ 62.45 & 7.93 & 83.55 & 5.61 \\
2008 Aug 20 & 18:30:21.68 & -23:06:39.7 & \ 10.0 & $-$6.0 & 105 & 104 & \ \ 12.49 & \ \ 60.42 & 8.61 & 87.81 & 6.09 \\
2008 Aug 21 & 18:02:53.91 & -34:31:17.8 & $-$3.0 & $-$6.0 & 111 & 109 & $-$25.85 & $-$29.81 & 8.33 & 86.94 & 5.89 \\
2008 Aug 21 & 18:07:17.57 & -32:47:21.0 & $-$1.0 & $-$6.0 & 111 & 110 & $-$11.74 & $-$7.74 & 9.26 & 97.15 & 6.55 \\
2008 Aug 21 & 18:11:48.68 & -31:01:50.7 & \ \ 1.0 & $-$6.0 & 114 & 112  &$-$21.54 & $-$9.42 & 9.16 & 96.93 & 6.48 \\
2008 Aug 21 & 18:16:10.33 & -29:19:15.1 & \ \ 3.0 & $-$6.0 & 115 & 114 & \ \ 10.17 & \ \ 30.18 & 8.98 & 95.91 & 6.35 \\
2008 Aug 22 & 18:00:21.58 & -35:24:53.8 & $-$4.0 & $-$6.0 & 114 & 112 & $-$43.53 & $-$51.61 & 8.35 & 88.41 & 5.91 \\
2008 Aug 22 & 18:05:10.43 & -33:40:05.1 & $-$2.0 & $-$6.0 & 112 & 109 & $-$40.98 & $-$40.96 & 8.55 & 89.29 & 6.05 \\
2008 Aug 22 & 18:20:20.14 & -27:33:33.3 & \ \ 5.0 & $-$6.0 & 111 & 109 & \ \ 25.20 & \ \ 53.17 & 8.46 & 88.38 & 5.99 \\
2008 Aug 23 & 18:09:46.09 & -31:56:01.1 & \ \ 0.0 & $-$6.0 & 110 & 109 & \ $-$8.53 & $-$0.42 & 8.70 & 90.82 & 6.15 \\
2008 Aug 24 & 18:13:54.50 & -30:07:05.2 & \ \ 2.0 & $-$6.0 & 112 & 110 & \ \ 4.04 & \ \ 20.30 & 8.31 & 87.11 & 5.87 \\
2008 Aug 24 & 18:18:25.27 & -28:25:38.5 & \ \ 4.0 & $-$6.0 & 116 & 113 & \ \ 4.99 & \ \ 29.13 & 8.18 & 87.00 & 5.79 \\
2008 Aug 24 & 18:24:29.51 & -25:47:41.7 & \ \ 7.0 & $-$6.0 & 111 & 111 & \ \ 19.66 & \ \ 55.65 & 8.36 & 88.10 & 5.91 \\
2008 Aug 24 & 18:26:51.06 & -20:27:05.7 & \ 12.0 & $-$4.0 & 105 & 104 & \ \ 29.39 & \ \ 85.59 & 7.84 & 79.94 & 5.54 \\
2008 Aug 25 & 18:30:38.75 & -18:36:13.3 & \ 14.0 & $-$4.0 & 110 & 110 & \ \ 34.63 & \ \ 98.83 & 7.65 & 80.26 & 5.41 \\
2008 Aug 25 & 18:38:04.33 & -15:04:28.9 & \ 18.0 & $-$4.0 & 106 & 103 & \ \ 28.39 & 107.73 & 6.09 & 61.77 & 4.30 \\
2008 Aug 25 & 18:45:38.33 & -11:32:11.6 & \ 22.0 & $-$4.0 & 103 & 101 & \ \ 37.81 & 132.19 & 6.36 & 63.91 & 4.50 \\
\hline
\end{tabular}
\end{scriptsize}
\end{table}
$^{a}$Number of velocities before the $\sigma$-clipping algorithm removed one star
in the calculation of the mean velocity and dispersion.
\clearpage

\begin{table}
\begin{scriptsize}
\centering
\caption{Data For Individual Stars in {\sl BRAVA} Fields}
\label{ind}
\begin{tabular}{p{0.4in}p{0.3in}p{0.3in}p{0.4in}p{0.4in}p{0.25in}p{0.25in}p{0.25in}p{0.42in}p{0.37in}p{0.25in}p{0.25in}p{0.25in}p{0.4in}p{0.4in}p{0.4in}} \\ \hline
Aperature & gal $l$ (deg) & gal $b$ (deg) & R.A. (J2000.0) & Decl. (J2000.0) & $J$ (mag) & $H$ (mag) & $K$ (mag) & $V_{HC}$   (km s$^{-1}$) & E(B-V) & $J_0$ (mag) & $H_0$ (mag) & $K_0$ (mag) & TiO(mag) & Spectrum (fits) & 2MASS ID\\ 
\hline
1 & 9.0256 & $-$7.8183 & 278.9434 & $-$24.7927 & 10.640 & 9.940 & 9.724 & $-$56.4 & 0.405 & 10.275 & 9.707 & 9.575 & $-$0.0345 & \underline{Spectrum} & 18354640-2447338 \\
3 & 8.6962 & $-$8.0373 & 279.0005 & $-$25.1823 & 11.116 & 10.381 & 10.136 & 101.8 & 0.418 & 10.739 & 10.140 & 9.983  & $-$0.0390 & \underline{Spectrum} & 18360011-2510562 \\
4 & 9.0018  & $-$7.8323  & 278.9456  & $-$24.8201  & 10.468  & 9.625  & 9.298  &	$-$84.1 & 0.403 & 10.105  & 9.393  &	9.150  & \ \  0.0055 & \underline{Spectrum}  & 18354694-2449123\\
7 & 9.2062  & $-$7.7561 & 278.9694  & $-$24.6047  & 10.643  & 9.843  & 9.590  &	$-$40.8 & 0.382 & 10.298 & 9.623 & 	9.450 & $-$0.0336 & \underline{Spectrum} & 18355264-2436170 \\
8  & 8.9748 & $-$7.9853  &	279.0837  & $-$24.9118  & 10.018 & 9.043 & 8.722 & 115.9 & 0.418 & 9.641 & 8.802 & 8.569 & \ \  0.0895 & \underline{Spectrum} & 18362009-2454424 \\
\hline
\end{tabular}
\end{scriptsize}
\end{table}

\begin{table}
\begin{scriptsize}
\centering
\caption{{\sl BRAVA} Rotation and Dispersion Results}
\label{normality}
\begin{tabular}{p{0.24in}p{0.24in}p{0.4in}p{0.35in}p{0.25in}p{0.35in}p{0.25in}p{0.45in}p{0.45in}} \\ \hline
gal $l$ (deg) & gal $b$ (deg) & Number Velocities & Skewness & $\sigma_{Skew}$ & Kurtosis & $\sigma_{kurt}$ & $p$-index (Shapiro-Wilk) & $p$-index (Anderson-Darling) \\ 
\hline
$-$6.5 & $-$3.1 & 63 & $-$0.60 & 0.31 & 0.07 & 0.62 & 0.043 & 0.126 \\ 
\ \ \ 5.5 & $-$3.1 & 86 & 0.05 & 0.26 & $-$0.31 & 0.53 & 0.842 & 0.991 \\ 
\ \ \ 0.0 & $-$3.5 & 111 & 0.19 & 0.23 & $-$0.41 & 0.46 & 0.035 & 0.004 \\
\ \ \ 4.0 & $-$3.5 & 106 & 0.22 & 0.24 & $-$0.64 & 0.48 & 0.182 & 0.266 \\
\ \ \ 1.0 & $-$4.0 & 61 & 0.14 & 0.31 & 0.86 & 0.63 & 0.238 & 0.267 \\
\ \ \ 7.0 & $-$4.0 & 93 & $-$0.08 & 0.25 & 0.37 & 0.51 & 0.386 & 0.502 \\
\ \ \ 6.0 & $-$4.0 & 106 & $-$0.11 & 0.24 & 0.09 & 0.48 & 0.967 & 0.943 \\ 
\ 12.0 & $-$4.0 & 104 & 0.07 & 0.24 & $-$0.71 & 0.48 & 0.203 & 0.412 \\ 
\ 14.0 & $-$4.0 & 110 & 0.28 & 0.23 & $-$0.23 & 0.47 & 0.500 & 0.306 \\ 
\ 18.0 & $-$4.0 & 103 & 0.27 & 0.24 & $-$0.38 & 0.48 & 0.434 & 0.498 \\ 
\ 22.0 & $-$4.0 & 101 & $-$0.38 & 0.24 & 0.72 & 0.49 & 0.258 & 0.567 \\ 
$-$1.0 & $-$4.3 & 109 & $-$0.17 & 0.23 & $-$0.09 & 0.47 & 0.513 & 0.160 \\
\ \ \ 4.0 & $-$4.5 & 108 & $-$0.05 & 0.24 & 0.02 & 0.47 & 0.750 & 0.384 \\
\ \ \ 6.0 & $-$4.8 & 106 & 0.26 & 0.24 & $-$0.37 & 0.48 & 0.357 & 0.747 \\ 
$-$6.5 & $-$5.1 & 79 & 0.22 & 0.28 & $-$0.06 & 0.55 & 0.789 & 0.728 \\ 
$-$10.0 & $-$6.0 & 103 & 0.15 & 0.24 & $-$0.36 & 0.48 & 0.440 & 0.163 \\ 
$-$9.0 & $-$6.0 & 102 & $-$0.02 & 0.24 & $-$0.50 & 0.49 & 0.899 & 0.883 \\ 
$-$8.0 & $-$6.0 & 105 & $-$0.24 & 0.24 & $-$0.26 & 0.48 & 0.449 & 0.762 \\ 
$-$7.0 & $-$6.0 & 103 & $-$0.04 & 0.24 & 1.96 & 0.48 & 0.087 & 0.495 \\ 
$-$6.0 & $-$6.0 & 96 & 0.43 & 0.25 & 1.43 & 0.50 & 0.116 & 0.542 \\ 
$-$5.0 & $-$6.0 & 108 & 0.25 & 0.24 & 0.57 & 0.47 & 0.075 & 0.329 \\ 
$-$4.0 & $-$6.0 & 112 & 0.38 & 0.23 & 0.11 & 0.46 & 0.084 & 0.426 \\ 
$-$3.0 & $-$6.0 & 109 & 0.01 & 0.23 & 0.55 & 0.47 & 0.803 & 0.506 \\ 
$-$2.0 & $-$6.0 & 109 & 0.06 & 0.23 & 1.62 & 0.47 & 0.130 & 0.364 \\ 
$-$1.0 & $-$6.0 & 110 & $-$0.18 & 0.23 & 0.31 & 0.47 & 0.898 & 0.835 \\ 
\ \ \ 0.0 & $-$6.0 & 109 & 0.09 & 0.23 & $-$0.62 & 0.47 & 0.360 & 0.298 \\ 
\ \ \ 1.0 & $-$6.0 & 112 & 0.22 & 0.23 & 0.03 & 0.46 & 0.793 & 0.614 \\ 
\ \ \ 2.0 & $-$6.0 & 110 & 0.26 & 0.23 & $-$0.36 & 0.47 & 0.517 & 0.754 \\ 
\ \ \ 3.0 & $-$6.0 & 114 & 0.24 & 0.23 & $-$0.55 & 0.46 & 0.092 & 0.109 \\ 
\ \ \ 4.0 & $-$6.0 & 113 & $-$0.33 & 0.23 & 0.60 & 0.46 & 0.083 & 0.556 \\ 
\ \ \ 5.0 & $-$6.0 & 109 & 0.18 & 0.23 & $-$0.58 & 0.47 & 0.436 & 0.523 \\ 
\ \ \ 6.0 & $-$6.0 & 102 & $-$0.35 & 0.24 & $-$0.09 & 0.49 & 0.104 & 0.121 \\ 
\ \ \ 7.0 & $-$6.0 & 111 & $-$0.03 & 0.23 & $-$0.37 & 0.46 & 0.661 & 0.507 \\ 
\ \ \ 8.0 & $-$6.0 & 107 & 0.31 & 0.24 & 0.14 & 0.47 & 0.572 & 0.459 \\ 
\ \ \ 9.0 & $-$6.0 & 111 & $-$0.48 & 0.23 & 0.03 & 0.46 & 0.680 & 0.186 \\ 
\ 10.0 & $-$6.0 & 192 & $-$0.19 & 0.18 & 0.79 & 0.35 & 0.153 & 0.342 \\ 
$-$6.0 & $-$6.9 & 97 & $-$0.12 & 0.25 & $-$0.80 & 0.50 & 0.121 & 0.292 \\ 
\ \ \ 6.0 & $-$6.9 & 98 & 0.32 & 0.25 & $-$0.22 & 0.49 & 0.133 & 0.110 \\ 
$-$10.0 & $-$8.0 & 97 & 0.27 & 0.25 & 0.07 & 0.50 & 0.622 & 0.443 \\ 
$-$9.0 & $-$8.0 & 104 & 0.00 & 0.24 & $-$0.01 & 0.48 & 0.894 & 0.864 \\ 
$-$8.0 & $-$8.0 & 99 & 0.24 & 0.25 & 0.53 & 0.49 & 0.655 & 0.545 \\ 
$-$7.0 & $-$8.0 & 102 & 0.41 & 0.24 & 1.01 & 0.49 & 0.059 & 0.117 \\ 
$-$6.0 & $-$8.0 & 94 & 1.38 & 0.25 & 4.91 & 0.51 & 0.0001 & 0.0001 \\ 
$-$5.0 & $-$8.0 & 108 & $-$0.06 & 0.24 & $-$0.10 & 0.47 & 0.627 & 0.649 \\ 
$-$4.0 & $-$8.0 & 99 & 0.78 & 0.25 & 2.68 & 0.49 & 0.002 & 0.227 \\
$-$3.0 & $-$8.0 & 114 & 0.01 & 0.23 & $-$0.32 & 0.46 & 0.466 & 0.509 \\ 
$-$2.0 & $-$8.0 & 102 & $-$0.42 & 0.24 & 0.16 & 0.49 & 0.255 & 0.320 \\
$-$1.0 & $-$8.0 & 96 & 0.71 & 0.25 & 2.42 & 0.50 & 0.006 & 0.079 \\
\ \ \ 0.0 & $-$8.0 & 102 & $-$0.01 & 0.24 & 0.03 & 0.49 & 0.922 & 0.773 \\
\ \ \ 1.0 & $-$8.0 & 103 & $-$0.05 & 0.24 & 0.40 & 0.48 & 0.762 & 0.686 \\
\ \ \ 2.0 & $-$8.0 & 103 & 0.21 & 0.24 & $-$0.09 & 0.48 & 0.467 & 0.268 \\ 
\ \ \ 3.0 & $-$8.0 & 108 & $-$0.33 & 0.24 & 0.89 & 0.47 & 0.422 & 0.299 \\ 
\ \ \ 4.0 & $-$8.0 & 104 & 0.11 & 0.24 & $-$0.13 & 0.48 & 0.802 & 0.934 \\ 
\ \ \ 5.0 & $-$8.0 & 102 & 0.24 & 0.24 & $-$0.05 & 0.49 & 0.519 & 0.481 \\ 
\ \ \ 6.0 & $-$8.0 & 98 & 0.43 & 0.25 & 0.29 & 0.49 & 0.173 & 0.110 \\
\ \ \ 7.0 & $-$8.0 & 102 & $-$0.23 & 0.24 & 0.60 & 0.49 & 0.169 & 0.182 \\ 
\ \ \ 8.0 & $-$8.0 & 99 & $-$0.89 & 0.25 & 4.16 & 0.49 & 0.0001 & 0.090 \\
\ \ \ 9.0 & $-$8.0 & 108 & $-$0.61 & 0.24 & 1.89 & 0.47 & 0.007 & 0.288 \\ 
\ 10.0 & $-$8.0 & 96 & 0.06 & 0.25 & $-$0.61 & 0.50 & 0.460 & 0.585 \\
\hline
\end{tabular}
\end{scriptsize}
\end{table}
\clearpage


\begin{thebibliography}{}
%
%
\bibitem[Babusiaux et~al.(2010)]{babusiaux10} 
Babusiaux, C., et al., 2010, A\&A, 519, 77
%
\bibitem[Beaulieu et~al.(2000)]{beaulieu00} Beaulieu, S. F., Freeman, K. C., Kalnajs, A. J., 
Saha, P., \& Zhao, H. 2000, AJ, 120, 855
%
\bibitem[Bissantz \& Gerhard(2002)]{bissantz02} Bissantz, N. \& Gerhard, O. 2002, MNRAS, 330, 59
%
\bibitem[Bekki \& Tsujimoto(2011)]{bekki11} Bekki, K. \& Tsujimoto, T.\ 2011, MNRAS.tmpL.288B
%
\bibitem[Blitz \& Spergel(1991)]{blitz91} Blitz, L. \& Spergel, D.N.\ 1991, ApJ, 370, 205
%
\bibitem[Brown et~al.(2005)]{brown05} Brown, W. R., Geller, M. J., Kenyon, S. J., \& Kurtz, M. J. 2005, ApJ, 622, L33
%
\bibitem[Cabrera-Lavers et~al.(2007)]{cabrera07} 
Cabrera-Lavers, A., Hammersley, P. L., Gonz$\rm \acute{a}$lez-Fern$\rm \acute{a}$ndez, C., 
L$\rm \acute{o}$pez-Corredoira, M., Garz$\rm \acute{o}$n, F., Mahoney, T. J.\ 2007, A\&A, 465, 825
%
\bibitem[Cavichia, Costa \& Maciel(2011)]{cavichia11} Cavichia, O., Costa, R. D. D. \& Maciel, W. J.\
MxAA, 47, 49
%
\bibitem[D'Agostino(1986)]{dagostino86} D'Agostino, R. B. 1986, in Goodness-of-Fit Techniques, 
ed. R. B. D'Agostino \& M. A. Stephens (New York: Dekker), 367
%
\bibitem[De Propris et~al.(2011)]{depropris11} De Propris, R. et al. 2011, ApJ, 732, 36
%
\bibitem[Dwek et~al.(1995)]{dwek95} Dwek, E. et al. 1995, ApJ, 445, 716
%
\bibitem[Fisher \& Drory(2011)]{fisher11} Fisher, D.B. \& Drory, N.\ 2011, ApJ, 733 L47
%
\bibitem[Frogel \& Whitford(1987)]{frogel87} Frogel, J.A. \& Whitford, A. E.\ 1987, ApJ, 320, L199
%
\bibitem[Frogel et~al.(1999)]{frogel99} Frogel, J.A., Tiede, G.P. \& Kuchinski, L. E.\ 1999, AJ, 117, 2296
%
\bibitem[Fulbright et~al.(2006)]{fulbright06} Fulbright, J.~P., McWilliam, A., \& Rich, R.~M.\ 2006, \apj, 636, 821 
%
\bibitem[Fulbright et~al.(2007)]{fulbright07} Fulbright, J.~P., McWilliam, A., \& Rich, R.~M.\ 2007, \apj, 661, 1152 
%
\bibitem[Gonzalez et~al.(2011)]{gonzalez11} Gonzalez, O.~A. et al. 2011, A\&A, 530, 54
%
\bibitem[Hammer et~al.(2007)]{hammer07} Hammer, F., Puech, M., Chemin, L., Flores, H.
\& Lehnert, M. D. 2007, ApJ, 662, 322
%
\bibitem[Hammersley et~al.(2000)]{hammersley00}
Hammersley, P. L., Garz\'{o}n, F., Mahoney, T. J., L\'{o}pez-Corredoira, M., \& Torres, M. A. P. 2000, MNRAS, 317, L45
%
\bibitem[Hauschildt et~al.(1999)]{hauschildt99} Hauschildt, P.~H., Allard, F., \& Baron, E.\ 1999, \apj, 512, 377 
%
\bibitem[Hauschildt et~al.(2003)]{hauschildt03} Hauschildt, P.~H., Allard, F., Baron, E., Aufdenberg, J., \& Schweitzer, A.\ 2003, ASP Conf. Ser., 298, 179 
%
\bibitem[Helmi et~al.(1999)]{helmi99}  Helmi, A., White, S. D. M., de Zeeuw, P. T., \& Zhao, H. 
1999, Nature, 402, 53
%
\bibitem[Hodge(1983)]{hodge83} Hodge, P. W. 1983, PASP, 95, 721
%
\bibitem[Howard et~al.(2008)]{howard08} Howard, C. D., Rich, R. M., Reitzel, D. B., Koch, A., 
De Propris, R., \& Zhao, H. 2008, ApJ, 688, 1060
%
\bibitem[Howard et~al.(2009)]{howard09} Howard, C. D. et al. 2009, ApJ, 702, L153
%
\bibitem[Jarvis(1990)]{jarvis90}
Jarvis, B. 1990, Dynamics and Interactions of Galaxies, ed. R. Wielen (New York: Springer), 416
%
\bibitem[Johnson et~al.(2011)]{johnson11} Johnson, C.~I., Rich, R.~M., Fulbright, J.~P., 
Valenti, E. \& McWilliam, A.\ 2011, ApJ, 732, 108
%
\bibitem[Kenyon et~al.(2008)]{kenyon08}
Kenyon, S. J., Bromley, B. C., Geller, M. J., \& Brown, W. R. 2008, ApJ, 680, 312 
%
\bibitem[Kepley et~al.(2007)]{kepley07} Kepley, A.A. et~al. 2007, AJ, 134, 1579
%
\bibitem[Kormendy \& Illingworth(1982)]{kormendy82}
Kormendy, J. \& Illingworth, G. 1982, ApJ, 256, 460
%
\bibitem[Kormendy \& Kennicutt(2004)]{kormendy04}
Kormendy, J. \& Kennicutt, R.~C., 2004, ARA\&A, 42, 603
%
\bibitem[Kormendy \& Barentine(2010)]{kormendy10}
Kormendy, J. \& Barentine, J.~C., 2010, ApJ, 715, L176
%
\bibitem[Kormendy et~al.(2010)]{kormendy10n}
Kormendy, J., Drory, N., Bender, R. \& Cornell, M.~E., 2010, ApJ, 723, 54
%
\bibitem[Launhardt et~al.(2002)]{launhardt02} Launhardt, R., Zylka, R., \& Mezger, P. G. 2002, A\&A, 384, 112
%
\bibitem[Liszt \& Burton(1980)]{liszt80} Liszt, H. S. \& Burton, M. B. 1980, ApJ, 236, 779
%
\bibitem[Lopez-Corredoira et al.(2005)]{lopez05} Lopez-Corredoira, M., Cabrera-Lavers, A.
\& Gerhard, O. E. 2005, A\&A, 439, 107
%
%
\bibitem[Mel\'{e}ndez et~al.(2008)]{melendez08}
Mel\'{e}ndez, J., et al., 2008, A\&A, 484, L21
%
\bibitem[Milone \& Barbuy(1994)]{milone94} Milone, A., \& Barbuy, B.\ 1994, \aaps, 108, 449 
%
\bibitem[Minniti et~al.(1995)]{minniti95} Minniti D. et al., 1995, MNRAS, 277, 1293
%
\bibitem[Mould(1986)]{mould86}
Mould, J.R.\ 1986, Stellar Populations, A.Renzini and M.Tosi, eds.(Cambridge: Cambridge Univ. Press) p. 9.
%
\bibitem[Nassau \& Blanco(1958)]{nassau58}Nassau, J. J. \& Blanco, V. M.\ 1958, ApJ, 128, 46
%
\bibitem[Nishiyama et~al.(2005)]{nishiyama05} Nishiyama, S. et al. 2005, ApJ, 621, L105
%
\bibitem[Nishiyama et~al.(2006)]{nishiyama06} Nishiyama, S. et al. 2006, ApJ, 647, 1093
%
\bibitem[Pfenniger(1984)]{pfenniger84} Pfenniger, D. 1984, A\&A, 134, 373 
%
\bibitem[Pfenniger(1985)]{pfenniger85} Pfenniger, D. 1985, A\&A, 150, 112 
%
\bibitem[Pfenniger \& Norman(1990)]{pfenniger90} Pfenniger, D. \& Norman, C. 1990, ApJ, 363, 391 
%
\bibitem[Press, Flannery \& Teukolsky(1986)]{press86} Press, W. H., Flannery, B. P., \& Teukolsky, S. A. 1986, Numerical Recipes. The Art of Scientific Computing (Cambridge: Cambridge Univ. Press) 
%
\bibitem[Proctor et~al.(2000)]{proctor00} Proctor, R. N., Sansom, A. E., \& Reid, I. N. 2000, 
MNRAS, 311, 37
\bibitem[Rangwala \& Williams(2009)]{rangwala09} Rangwala, N., \& Williams, T.~B.\ 2009, \apj, 702, 414 
%
\bibitem[Rattenbury et~al.(2007)]{rattenbury07} Rattenbury, N. J., Mao, S., Sumi, T. \& Smith, M. C.
2007, MNRAS, 378, 1064	
%
\bibitem[Reitzel et~al.(2007)]{reitzel07} Reitzel, D. B., et al. 2007, BAAS, 39, 897 
%
\bibitem[Rich \& Origlia(2005)]{rich05} Rich, R.~M., \& Origlia, L.\ 2005, \apj, 634, 1293 
%
\bibitem[Rich et~al.(2007a)]{rich07a} Rich, R.~M., Origlia, L., \& Valenti, E.\ 2007, ApJ, 665, L119 
\bibitem[Rich et~al.(2007b)]{rich07b} Rich, R.~M., Reitzel,  D.~B., Howard, C.~D., \& Zhao, H.\ 2007, ApJ, 658, L29 
%
\bibitem[Robin et~al.(2011)]{robin11} Robin, A.C., Marshall, D.J., Schultheis, M. \& 
Reyle, C.\ et~al. 2011, 2011arXiv1111.5744 , accepted to A\&A
%
\bibitem[Rutledge et~al.(1997)]{rutledge97} Rutledge, G.~A., Hesser, J.~E., Stetson, P.~B., Mateo, M., Simard, L., Bolte, M., Friel, E.~D., \& Copin, Y.\ 1997, \pasp, 109, 883 
%
\bibitem[Schlegel et~al.(1998)]{schlegel98} Schlegel, D.J., Finkbeiner, D.P., \& Davis, M. 1998, \apj, 500, 525
%
\bibitem[Sevenster et~al.(1999)]{sevenster99} Sevenster, M., Saha, P., Valls-Gabaud, D. \& Fux, R.\ 1999,
MNRAS, 307, 584
%
\bibitem[Sharples et~al.(1990)]{sharples90} Sharples, R., Walker, A., \& Cropper, M.\ 1990, MNRAS, 246, 54 
%
\bibitem[Shapiro \& Wilk(1965)]{shapiro65} Shapiro, S. S., \& Wilk, M. B. 1965, Biometrika, 52, 591 
%
\bibitem[Shaw(1993)]{shaw93} Shaw, M. 1993, A\&A, 280, 33
%
\bibitem[Shen et~al.(2010)]{shen10} Shen, J., Rich, M.R., Kormendy, J., Howard, C., De Propris, R. \& Kunder, A.\
2010, ApJ, 720, L72
%
\bibitem[Skrutskie et~al.(2006)]{skrutskie06} Skrutskie, M. F. et al. 2006, AJ, 131, 1163
%
\bibitem[Stanek et~al.(1994)]{stanek94} Stanek, K. Z., Mateo, M., Udalski, A., Szymanski, M.,
Kaluzny, J. \& Kubiak, M. 1994, ApJ, 429, L72
%
\bibitem[Stanek et~al.(1997)]{stanek97} 	Stanek, K. Z., Udalski, A., Szymanski, M., Kaluzny, J.,
Kubiak, M., Mateo, M. \& Krzeminski, W. 1997, ApJ, 477, 163
%
\bibitem[Sumi et~al.(2004)]{sumi04} Sumi, T.\ 2004, MNRAS, 348, 1439
%
\bibitem[Terndrup et~al.(1990)]{terndrup90} Terndrup, D.~M., Frogel, J.~A., \& Whitford, A.~E.\ 1990, \apj, 357, 453 
%
\bibitem[van Dokkum et~al.(2001)]{dokkum01} van Dokkum, P.G.\ 2001, PASP, 113, 1420
%
\bibitem[Weiland et~al.(1994)]{weiland94} Weiland, J. L. et al. 1994, ApJ, 425, L81
%
\bibitem[Zhao(1996)]{zhao96} Zhao, H.\ 1996, ASPC, 91, 549
%
\bibitem[Zoccali et~al.(2008)]{zoccali08} Zoccali, M. et al. 2008, A\&A, 486, 177
%
\end{thebibliography}
\end{document}